\documentclass[twocolumn,aps,epsfig,showpacs,floatfix]{revtex4}
\usepackage{epsfig}
\usepackage{dcolumn}
\usepackage{bm}
\usepackage{longtable}

\begin{document}


\title{Phobos results on charged particle multiplicity and pseudorapidity distributions in Au+Au, Cu+Cu, d+Au, and p+p collisions at ultra-relativistic energies}

\author{
B.Alver$^4$,
B.B.Back$^1$,
M.D.Baker$^2$,
M.Ballintijn$^4$,
D.S.Barton$^2$,
R.R.Betts$^6$,
A.A.Bickley$^7$,
R.Bindel$^7$,
A.Budzanowski$^3$, 
W.Busza$^{4,}${\protect \footnote{E-mail: busza@mit.edu {\bf Spokesperson}}}, 
A.Carroll$^2$,
Z.Chai$^2$,
V.Chetluru$^6$,
M.P.Decowski$^4$,
E.Garc\'{\i}a$^6$, 
T.Gburek$^3$,
N.George$^2$,
K.Gulbrandsen$^4$, 
S.Gushue$^2$,
C.Halliwell$^6$,
J.Hamblen$^8$, 
G.A.Heintzelman$^2$, 
C.Henderson$^4$,
D.J.Hofman$^6$,
R.S.Hollis$^6$, 
R.Ho\l y\'{n}ski$^3$,  
B.Holzman$^2$, 
A.Iordanova$^6$,
E.Johnson$^8$, 
J.L.Kane$^4$, 
J.Katzy$^4$,
N.Khan$^8$,
J.Kotu\l a$^3$
W.Kucewicz$^6$, 
P.Kulinich$^4$,
C.M.Kuo$^5$,
W.Li$^4$,
W.T.Lin$^5$, 
C.Loizides$^4$,
S.Manly$^8$,  
D.McLeod$^6$, 
J.Micha\l owski$^3$, 
A.C.Mignerey$^7$, 
R.Nouicer$^{2,6}$, 
A.Olszewski$^3$, 
R.Pak$^2$, 
I.C.Park$^8$,
H.Pernegger$^4$,
C.Reed$^4$, 
L.P.Remsberg$^2$, 
M.Reuter$^6$, 
C.Roland$^4$, 
G.Roland$^4$, 
L.Rosenberg$^4$,
J.Sagerer$^6$,
P.Sarin$^4$, 
P.Sawicki$^3$,
I.Sedykh$^2$,
W.Skulski$^8$, 
C.E.Smith$^6$,
S.G.Steadman$^4$, 
P.Steinberg$^2$,
G.S.F.Stephans$^4$, 
M.Stodulski$^3$,
A.Sukhanov$^2$, 
M.B.Tonjes$^7$,
A.Trzupek$^3$, 
C.Vale$^4$, 
G.J.van~Nieuwenhuizen$^4$, 
S.S.Vaurynovich$^4$,
R.Verdier$^4$, 
G.I.Veres$^4$,
B.Wadsworth$^4$, 
P.Walters$^8$,
E.Wenger$^4$,
F.L.H.Wolfs$^8$, 
B.Wosiek$^3$, 
K.Wo\'{z}niak$^3$, 
A.H.Wuosmaa$^1$, 
B.Wys\l ouch$^4$\\
\vspace{3mm}
\small
$^1$~Argonne National Laboratory, Argonne, IL 60439, USA\\
$^2$~Brookhaven National Laboratory, Upton, NY 11973, USA\\
$^3$~Institute of Nuclear Physics PAN, Krak\'{o}w, Poland\\
$^4$~Massachusetts Institute of Technology, Cambridge, MA 02139, USA\\
$^5$~National Central University, Chung-Li, Taiwan\\
$^6$~University of Illinois at Chicago, Chicago, IL 60607, USA\\
$^7$~University of Maryland, College Park, MD 20742, USA\\
$^8$~University of Rochester, Rochester, NY 14627, USA\\
}

Final version: \today

\begin{abstract}

Pseudorapidity distributions of charged particles emitted in $Au+Au$, $Cu+Cu$, $d+Au$, and $p+p$ collisions over a wide energy range have been measured using the PHOBOS detector at RHIC. The centrality dependence of both the charged particle distributions and the multiplicity at midrapidity were measured. Pseudorapidity distributions of charged particles emitted with $|\eta|<5.4$, which account for between 95\% and 99\% of the total charged-particle emission associated with collision participants, are presented for different collision centralities. Both the midrapidity density, $dN_{ch}/d\eta$, and the total charged-particle multiplicity, $N_{ch}$, are found to factorize into a product of independent functions of collision energy, $\sqrt{s_{_{NN}}}$, and centrality given in terms of the number of nucleons participating in the collision, $N_{part}$. The total charged particle multiplicity, observed in these experiments and those at lower energies, assumes a linear dependence of $(\ln s_{_{NN}})^2$ over the full range of collision energy of $\sqrt{s_{_{NN}}}$=2.7-200 GeV.    
 
\end{abstract}

\vspace{0.5cm}
\maketitle

\section{Introduction}

The study of relativistic heavy ion collisions is the only known method of creating and studying in the laboratory systems with hadronic or partonic degrees of freedom at extreme energy and matter density over a significant volume.  It is for this reason that in recent years such studies have attracted much experimental and theoretical interest, in particular with the likelihood that at the higher energies a new state of QCD matter is created. 

During the first five years of the operation of the Relativistic Heavy Ion Collider, RHIC, at Brookhaven National Laboratory, the PHOBOS experiment~\cite{PhobosNIM} collected extensive data on the production of charged particles over almost the entire solid angle, for a wide range of collision energies and colliding nuclei. Many interesting and unexpected results were obtained which have been published and their significance discussed in a series of short papers \cite{PH1,PH3,PH4,PH5,PH7,PH10,PH15,PH17,PH23,PH25,PH29,PH34}. The early results are summarized and the physics interpretation is discussed in Ref.~\cite{PH24}. 

This paper presents all PHOBOS results on multiplicity and pseudorapidity distributions, including some unpublished data, in a consistent graphical and tabular form, together with detailed descriptions of how the results were obtained and analyzed. The intention is to present the data with a minimum of interpretation. Fitting of functional forms to the data is done only to facilitate reproduction or extrapolation. No significance of the functional forms is implied. 

The PHOBOS data cover {\it Au+Au} collisions at nucleon-nucleon center of mass energy, $\sqrt{s_{_{NN}}}$, of 19.6, 56, 62.4, 130 and 200 GeV, {\it Cu+Cu} at 22.4, 62.4 and 200 GeV, {\it d+Au} at 200 GeV, and {\it p+p} at 200 and 410 GeV. Similar measurements, though with less extensive coverage, have been made by the other RHIC experiments BRAHMS~\cite{BRAHMSmult}, STAR~\cite{STARmult}, and PHENIX~\cite{PHENIXmult}. These measurements extend earlier studies of $p+A$ collisions at Fermilab \cite{refA,refB}, $p+A$ collisions at the Super Proton Synchrotron (SPS) at CERN \cite{refC}, $p+Nuclear Emulsion$ \cite{refD}, as well as A+A collisions at the SPS reaching energies up to $\sqrt{s_{_{NN}}}$ = 17.3~GeV \cite{NA49}, and at the Alternating Gradient Synchrotron (AGS) at BNL up to 4.9 GeV \cite{Klay}. It is expected that heavy-ion collisions will soon be extended to higher energies, eventually reaching $\sqrt{s_{_{NN}}}$ = 5500~GeV at the Large Hadron Collider at CERN.

This extensive body of data on the global properties of particle production in heavy ion collisions can be used to provide insight into both our understanding of the mechanisms of particle production and the properties of matter that exist at extremes of energy and matter densities.

This paper is organized as follows: The PHOBOS apparatus relevant for the multiplicity measurements is briefly described in Sect. II. This is followed in Sect. III by a detailed discussion of the data analysis procedure. The results are presented in Sect. IV,  and a summary is given in Sect. V.

\section{Experimental Setup}
\label{ExpSetup}

\begin{figure}
\epsfig{file=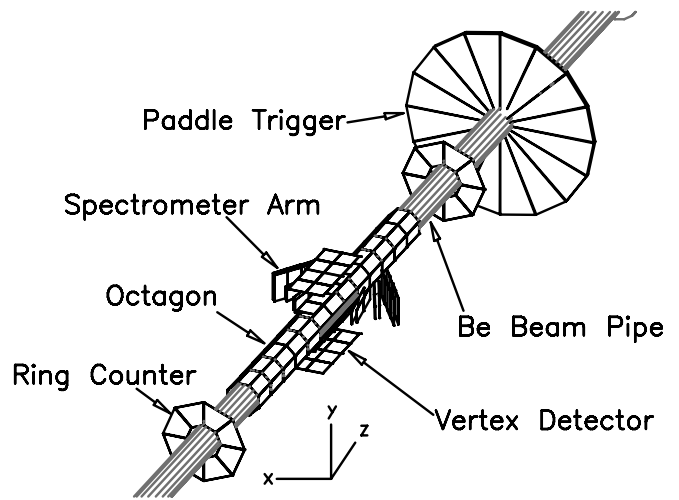,width=\columnwidth}
\vspace{2mm}
\caption{The position and orientation of the Be beam pipe and the active areas of several of the detectors used in the present work. See text for details.}
\label{Phmult_fig1}
\end{figure}

The PHOBOS experiment consists of three major components, a charged particle multiplicity detector covering a large fraction of the total solid angle, an array of detectors for triggering and event characterization, and a two-arm magnetic spectrometer used for reconstructing the trajectories of a small fraction of the particles emitted near midrapidity. The entire detector is described in greater detail in Ref. \cite{PhobosNIM}. Note that only a sub-set of detectors were installed for the run resulting in the data presented in Ref.~\cite{PH1}. This section will briefly discuss the parts of the apparatus used in the current analysis. The active areas of several of the detectors used in this work as well as the beam pipe are shown in Fig. \ref{Phmult_fig1}. Note that the dimensions of all detectors as well as the positions transverse to the beam are shown to scale; the locations along the beam have been shifted to facilitate the viewing of the detectors in a single figure. The Paddle counter array on one side of the interaction point and the outer four Ring counters are not shown. The dimensions and orientations of the excluded detectors are identical to those shown in Fig.~\ref{Phmult_fig1}.

The event triggering and centrality determination were, for most of the systems, provided by the Paddle detector, two arrays of 16 plastic scintillator slats positioned at $\pm$ 3.21 m from the center of the interaction region~\cite{PaddleCounters}. Each slat is read out by a single photomultiplier tube connected to the outer end by a light guide (not shown). The active area of the Paddle detectors covers the angular region $3 < |\eta|< 4.5$, where $\eta=-\ln[\tan(\theta/2)]$ and $\theta$ is the polar angle defined with respect to the beam axis $z$. The primary event trigger required response from at least one slat in both counters with a time difference consistent with an event occuring near the center of the interaction region. Detailed analysis and comparison to simulations indicate that this trigger fired for $>97\%$ of the Au+Au total nuclear cross-section at $\sqrt{s_{_{NN}}}$= 130 and 200 GeV, and $\sim$81\% for the $\sqrt{s_{_{NN}}}$ = 19.6 GeV data \cite{Hollis_Bari}, whereas the trigger efficiency for $Cu+Cu$ varies between 84\% for 200 GeV and 75\% for the 62.4 GeV collisions~\cite{PH34}. The same trigger conditions were required for the 200 GeV $d+Au$ experiment, resulting in an overall triggering efficiency of $\sim$83\%, whereas the inelastic $p+p$ collisions were obtained by requiring only one slat in one counter to trigger in coincidence with the signal from the beam bunch crossing clock provided by RHIC \cite{Sagerer_thesis}.

The Vertex detector was used in both event characterization and multiplicity determination. It consists of four layers of Si (silicon) wafers, two above and two below the interaction region. This detector covers the two regions, each with an azimuthal, $\phi$, angular extent of roughly 43$^\circ$ and $\eta$ range (for events occuring at the center of the interaction 
region) of $|\eta| \leq 1$. The Si detectors are finely segmented along the beam direction so that ``tracklets'', created by combining one hit from the inner and one hit from the outer layers, pointed back to the primary interaction point with high accuracy. This vertex location was then used to correct the signal in other parts of the multiplicity detector (especially the Octagon) for the effect of traversing the Si wafers at oblique angles. In addition, the distribution of tracklets was used to measure the charged particle multiplicity near midrapidity. 

The primary multiplicity detectors are the Octagon and the Rings. The 
former consists of a single layer of 92 Si wafers oriented parallel to the beam pipe and covering $|\eta|<3.2$. Except for regions left open to allow unimpeded passage to the Vertex and Spectrometer detectors, the Octagon has full azimuthal ($\phi$) coverage. The wafers are segmented in both $\phi$ (about 10$^\circ$) and $\eta$ (ranging from 0.06 to 0.005 units depending on distance from the center). The Rings consist of six separate detector arrays (only two are shown in Fig.~\ref{Phmult_fig1}) located at $\pm$1.13 m, $\pm$2.35 m, and $\pm$5.05 m along the beam axis, extending the coverage for charged particle detection out to $|\eta| < 5.4$. These wafers (eight in each Ring) are oriented perpendicular to the beam and are segmented in both $\phi$ and $\eta$, with the radial segmentation chosen to give approximately constant $\Delta \eta$ bin sizes within a single detector. These detectors have full $\phi$ coverage.

\begin{figure}
\epsfig{file=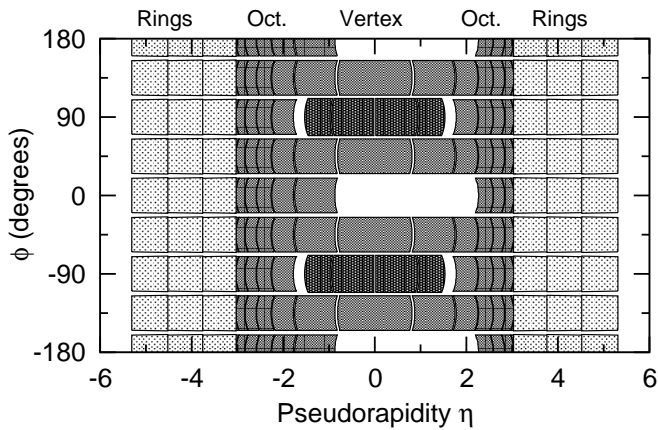,width=\columnwidth}

\caption{The geometrical acceptances of the Ring (light), Octagon (medium), and Vertex (dark) detectors are shown as a function of pseudorapidity, $\eta$, and azimuthal angle, $\phi$, for particles emitted from the nominal interaction point at the center of the Octagon array.
}
\label{acceptance}
\end{figure}

The geometrical acceptance of the detectors used in the charged particle multiplicity measurements is shown in Fig.~\ref{acceptance}. The pseudorapidity range is here calculated assuming that collisions occur at the center of the octagon array. The openings at $\phi$ = 0$^\circ$ and 180$^\circ$ and -0.8$<\eta<$2.1 are partially covered by the Two-arm spectrometer (not shown). Selected characteristics of the Si wafers are listed in Table~\ref{Si_wafers}. The pad dimensions in the longitudinal direction are for all Si-sensors sufficient to sort the data into $\eta$-bins of width $\Delta \eta <$ 0.2, which is the bin-size used for the multiplicity distributions presented in this work.

\begin{table}
\caption{Geometrical characteristics of Si sensors used in charged particle multiplicity measurements. All Si wafers have thicknesses in the range 300-340 $\mu$m.}
\vspace{3mm}
\begin{tabular}{lccc}
\tableline
Sensor type	& Active area  		& Number of pads 	& Pad size	\\
		& (mm$^2$)		& $\eta \times  \phi$		& (mm$^2$)     \\ 	
\hline
Octagon		& 81.28 $\times$ 34.88	& 30 $\times$4 		& 2.71$\times$8.71  	\\
Ring		& ${\rm{\approx}}$3200	& 8 $\times$ 8     	& ${\rm{\approx}}$ 20-105 \\
Inner Vtx	& 60.58 $\times$ 48.18 	& 256 $\times$ 4 	& 0.47 $\times$ 12.04	\\
Outer Vtx	& 60.58 $\times$ 48.18 	& 256 $\times$ 2 	& 0.47 $\times$ 24.07 	\\
\tableline
\end{tabular}
\label{Si_wafers}
\end{table}

The Si wafers of the Spectrometer detector were designed primarily to track and identify charged particles emitted near $\eta=0$. As a secondary function, the six planes outside the Spectrometer magnetic field, shown in Fig. \ref{Phmult_fig1}, are also used to measure the charged particle multiplicity within the range $|\Delta \eta|<1$. Straight-line tracks formed by aligning hits from each of the six layers were used in this analysis. 

In studying the charged particle multiplicity, it is critical that the experimental setup minimizes the modifications of the distribution due to absorption, scattering, or creation of secondaries. The Be beam pipe is an important element of the PHOBOS design. It is constructed of three identical segments, each 4 m long by 
76 mm in diameter, with a wall thickness of $\approx$ 1 mm. The flanges and bolts used to connect the segments are also made of Be. The use of a low-Z material and relatively thin walls reduces the interactions of charged particles traversing the pipe. The relatively small diameter allows active detector elements to be positioned quite close to the interaction point as well as at small values of $\theta$ (and hence large $\eta$).

\section{Data analysis}
\label{DataAnalysis}

\subsection{Centrality determination}

In interpreting data from heavy ion collisions, the primary event characterization parameters are the energy of the collision and the overlap of the two nuclei at the moment when they interact, commonly referred to as centrality. Conventionally, $\sqrt{s_{_{NN}}}$, the center of mass energy available when a single nucleon from one projectile collides with a single nucleon from the other projectile (ignoring Fermi motion), is used to characterize the energy of the collision \cite{PH24}. 

Characterization of the centrality is more difficult since it is not a directly measurable quantity. Conventionally, three derived quantities are used as a measure of centrality. The quantity most closely related to measurement is the  so-called fractional cross-section. For each event, the observed energy or multiplicity of charged particles emitted into a fixed set of detectors is measured. Events are then sorted into bins of fractional total cross-section, making the reasonable assumption that the detector response has a monotonic relationship with centrality bins, with the bin containing events with the highest multiplicity corresponding to the most central collisions, {\it i.e.} those with the smallest impact parameter.

In PHOBOS, for the {\it Au+Au} data obtained at $\sqrt{s_{_{NN}}}$ = 56, 62.4, 130, and 200 GeV, and the {\it Cu+Cu} data at 62.4 and 200 GeV, it is the energy deposited in the Paddle detector arrays \cite{PaddleCounters} that is used for sorting the data into bins of fractional cross-section. Since the triggering efficiency plays an important role, especially for low-multiplicity peripheral events, it is important to accurately estimate this quantity. This is done by detailed comparisons with Monte Carlo simulations as described in detail in Ref.~\cite{Hollis_Bari}. The Paddle detectors cover the range 3$<|\eta|<$4.5, into which a large multiplicity of charged particles are emitted at these energies. For $\sqrt{s_{_{NN}}}$=19.6 GeV {\it Au+Au} and 22.4 GeV {\it Cu+Cu} collisions, however, the charged particle multiplicity is small in the large $|\eta|$ region covered by the trigger detectors, such that these centrality measurements become less reliable. At these lower energies a different measure, the total energy deposited in the Octagon detector in the range -3.0$<\eta<$3.0, is used for sorting the data. The centrality measurements for {\it d+Au} collisions represent a special challenge. For reasons discussed in Ref.{\cite{PH23}}, the six Ring detectors spanning 3.0$<|\eta|<$5.4 were used to obtain the most consistent centrality measurement.  

A detailed discussion of the choices of these detectors and of the fractional cross-section determination in PHOBOS can be found in Ref.~\cite{PH24}.

There are two other measures of centrality that have been found more useful when it comes to comparison of heavy-ion data with {\it p+p} data, as well as for the interpretation of the data, and which can be derived from the fractional cross-section. Both are motivated by the fact that, because of relativistic time dilation, the collision duration time at high energies is very short compared to the typical time-scale for soft particle production and for nuclear rearrangement or movement of nucleons within the nuclei. Assuming that the nucleons in each nucleus are indeed unaltered and fixed in the transverse direction during the time of the collision of the two nuclei, one can introduce the concept of $N_{coll}$, the number of individual nucleon-nucleon collisions that occur during the nucleus-nucleus collision. One can also introduce the concept of $N_{part}$, the number of nucleons that have made at least one collision with a nucleon of the other nucleus during the collision. The latter is the same quantity as the number of ``wounded nucleons'' introduced by Bialas {\it et al.}~\cite{Bialas} to interpret the {\it p+A} results obtained by Fermilab experiment E178 \cite{Busza}.

Both $N_{part}$ and $N_{coll}$ can be derived from the fractional cross-section by modeling the collisions of the two nuclei and assuming that the fractional cross-section increases monotonically with $N_{part}$ or $N_{coll}$. In PHOBOS, we model the collisions assuming the nuclei are a collection of hard spheres, with radii corresponding to the inelastic nucleon-nucleon cross-section at the appropriate energy, distributed according to the Wood-Saxon functional form for Cu and Au ions and the Hulth\'en wave function for deuterons, and that the nucleons are unaltered and follow straight-line trajectories as they make collisions inside the nucleus (a procedure often referred to as the Glauber model~\cite{Glauber}). Details of the modeling can be found in Ref~\cite{PH24}. 

\begin{figure}
\centerline{\epsfig{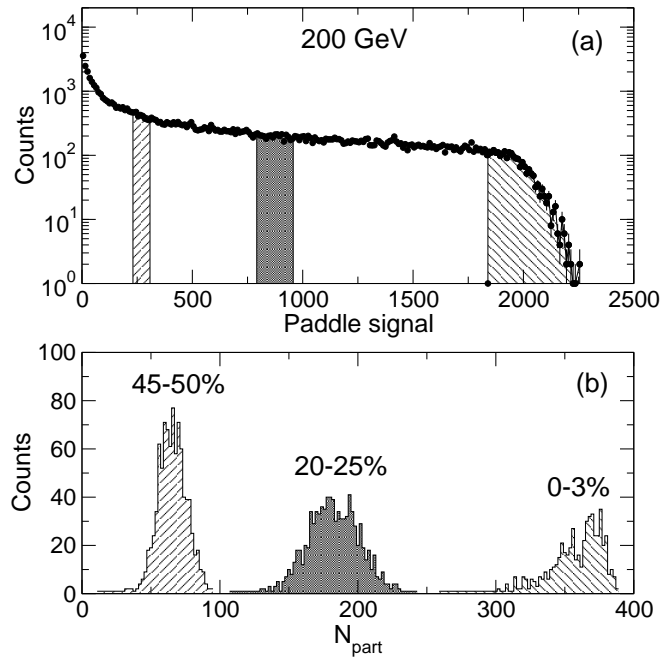}}
\caption{ Illustration of the conversion from paddle signal to $N_{\it part}$ for 200 GeV $Au+Au$ collisions. Panel a) shows the ranges of the Paddle signals corresponding to the 0-3\%, 20-25\%, and 45-50\% most central collision. The corresponding distributions of the number of participants, $N_{part}$, are shown in panel (b).}
\label{Phmult_fig3}
\end{figure}

Figure~\ref{Phmult_fig3} shows the total energy deposition spectrum from the Paddle detectors covering 3$<|\eta|<$4.5 as well as the deduced number of participants, $N_{\it part}$, for the three centrality bins of 0-3\%, 20-25\%, and 45-50\% of the total cross section in 200 GeV $Au+Au$ collisions (indicated by shaded bands in Fig.~\ref{Phmult_fig3}a). The corresponding $N_{\it part}$ distributions are shown in Fig.~\ref{Phmult_fig3}b and values of the centroids of these $N_{\it part}$ distributions associated with the cross section bins are listed in Table~\ref{table1} (see Appendix).  

\subsection{Signal processing}

Nine types of silicon pad sensors are used in PHOBOS. These sensors are segmented into as many as 1536 individual pads in the case of first spectrometer detector plane. Each pad is connected to one channel of a 128- or 64-channel pre-amplifier and readout chip. When a particle traverses the
detector a trigger-derived ``hold'' signal causes a front-end chip to
store the signal peak on all channels simultaneously and later
multiplexes them onto a differential analog output bus. 
The technical details of the Si-detectors and the associated electronics 
are described in Ref.~\cite{PhobosNIM} and references therein~\cite{RN,Ny,To,Calibration}.

For every event satisfying the trigger criteria,
the silicon data are converted from raw 12-bit ADC values to
calibrated deposited energies by applying a series of four algorithms, namely: 1) pedestal subtraction, 2) common mode noise correction, 3) gain correction, and 4) hit merging. 

The pedestals of the read-out system are subtracted from the raw signal data event-by-event for each channel. The pedestal widths provide a measure of the noise in the system that arises due to silicon detector leakage currents, as well as intrinsic electronics noise. The pedestals are obtained from dedicated runs with triggers derived from a random electronic pulser with adjustments using low multiplicity collision data.

\begin{figure}
\centerline{\epsfig{file=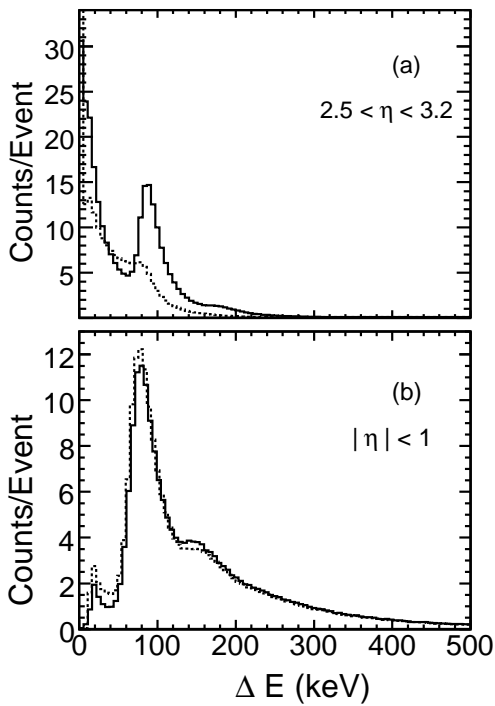,width=6.5cm}}
\caption{a) Energy deposition spectra for $2.5<\eta<3.2$ for the Octagon detector. b) Same as a) but for $|\eta|<1$. The dotted histogram corresponds to the raw spectrum, whereas the solid histogram shows the effect of the signal merging correction.}
\label{Phmult_fig4}
\end{figure}

The detector readout system also has a noise component that fluctuates with time, but is common to several channels. The level of this ``common mode'' noise (CMN) is
estimated on an event-by-event basis. The most probable value within a certain range of the pedestal subtracted ADC signals of all channels from one read-out chip is considered to represent the CMN for the chip. This value is subtracted from all channels of the relevant chip.

For very high occupancies, the data-based common mode noise correction in the octagon detector ($|\eta|<3.2$) 
becomes slightly inaccurate since it relies on having enough empty channels to provide a baseline. The more highly 
segmented vertex detector allows us to measure this effect and make a further data-based correction. This further 
correction was only required near midrapidity ($|eta|<1.5$) for the central (0-10\%), high energy 
($\sqrt{s_{_{NN}}}\ge 130 GeV$) Au+Au data, and it was less than 4\% everywhere.

The conversion from ADC channel to energy deposition in the Si-wafer is obtained by using a calibration circuit~\cite{Calibration}, which injects a known amount of charge into the front end of each channel during special calibration runs. We have observed that the gain of the front-end chips is linear over most of the range. Non-linearities are observed only for very large signals corresponding to the energy deposition of around 65 minimum ionizing particles (65 MIPs), where
1 MIP corresponds to about 80 keV most probable energy loss in 300 ${\rm \mu }$m silicon.

\begin{figure}
\centerline{\epsfig{file=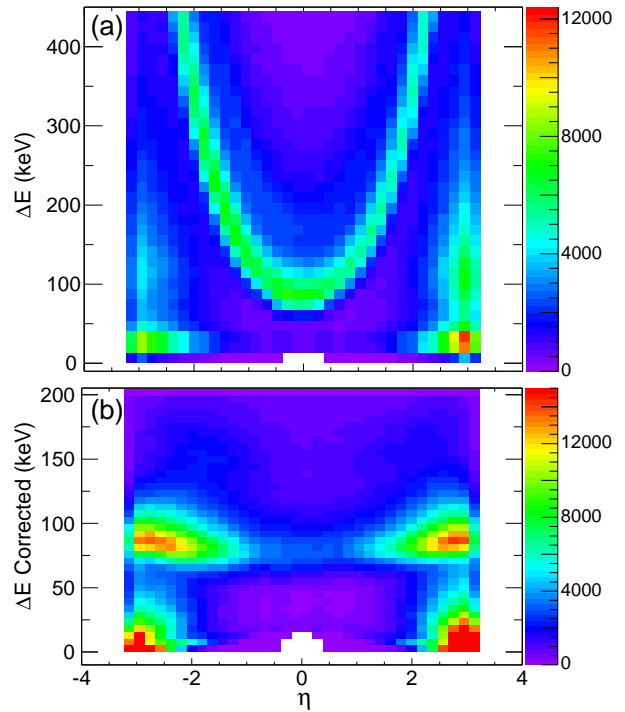,width=\columnwidth}}
\caption{(color online) (a) Map of the raw energy deposition versus pseudorapidity $\eta$ for the Octagon detector. (b) Same as (a) after applying the merging of signals from particle traversing multiple pads and correcting for the angle of incidence.}
\label{Phmult_fig5}
\end{figure}

In the center of the Octagon and in the Rings, primary particles were incident very close to normal to the silicon wafers (depending on the location of the primary interaction), and, therefore,
deposited an amount of energy corresponding to that of a minimum ionizing particle.  Near the Octagon extremities, a particle's trajectory is at very oblique angles, and the particle may  deposit energy in two or more adjacent pads.  In this case the deposited energy from all pads contributing to a single particle track was merged. The location of each pad with a
deposited energy above the hit threshold of $E_{th}$=19.2 keV was
identified.  For pads where the detector geometry indicates that the
particle track must traverse more than a single pad, the nearest
neighbor pads are checked to determine if they also contain some
deposited energy.  These signals are added together, and assigned a
pseudorapidity value appropriate for the pad that contained the
largest energy signal for that track.  The nearest-neighbor pads are
no longer considered in the subsequent analysis.  Figure
\ref{Phmult_fig4}a illustrates the effects of this merging procedure.  The dashed and solid histograms show deposited-energy spectra for unmerged and merged hits, respectively, over an $\eta$ range of 2.5 to 3.2.  The improvement in the signal quality is evident.  This correction is much smaller near midrapidity as illustrated in Fig.~\ref{Phmult_fig4}b, which corresponds to $|\eta|<1$. Note that these spectra have already been corrected for the angle-of-incidence as described below.

After merging hits, the deposited energy, $\Delta E_{tr}$, was corrected for its angle of incidence, as well as the thickness of the silicon wafer, so that the deposited energy could be compared to a common most probable value of $\Delta E_{\it MIP}$=82 keV. The uncorrected energy deposition (normalized to a common wafer thickness of 300 $\mu$m; the thickness of the Si wafers were measured and found to lie in the range 300-340~$\mu$m) is shown as a function of $\eta$ in Fig.\ref{Phmult_fig5}a.  For valid primary tracks, we observe that the energy deposition for a single minimum ionizing particle follows a $\Delta E \approx \cosh \eta$ dependence corresponding to the length of the track in the Si wafer.  Figure \ref{Phmult_fig5}b shows the angle-of-incidence corrected energy deposition, $\Delta E$, plotted as a function of $\eta$.  To select primary tracks created at the collision vertex, only tracks with $\Delta E$ values greater than 0.6$\times \Delta E_{\it MIP}$= 45 keV were
included in the subsequent analysis.  This selection eliminated a large fraction of tracks from secondary sources, $e.g.$ the beam-pipe, magnet yokes, etc., that have a small $\Delta E$ value inconsistent with primary particles.  Many such tracks are seen in the region of small $\Delta E$ values and large $|\eta|$ in Fig.\ref{Phmult_fig5}b. This background was insignificant for small values of $\eta$, where the amount of extra material between the interaction point and the detectors is minimal. 

Malfunctioning channels are flagged and stored in a
``dead channel'' map. Additional dead channels are found by looping over events in a reference run. When the signal in a channel was 5 times larger than its noise  it was considered a valid hit and a hit-counter was increased and the energy value
added to an energy sum. At the end of a run, the average energy  a particle deposited in a silicon pad was calculated and compared to a set of criteria, which included a minimum and maximum hit occupancy for the channel and a range of the average energy deposition per hit. Channels that failed to satisfy these criteria were marked ``dead'' in a map.

\subsection{Vertex finding}

At RHIC, the ion beams collide at zero degrees, such that the interaction vertices are distributed over a relatively large region of about 1 m along the beam-line around the nominal collision point. 
An approximate measurement ($\pm$7.5 cm) of the  $z$-coordinate of the collision vertex was obtained from the relative time difference between the paddle counters used in the event triggering. Also, the density distribution along the $z$-axis of hits with an energy deposition between 0.5 and 2 $\Delta E_{MIP}$ yielded a vertex finding  accuracy of $\Delta z\sim$ 0.5 cm. Although these detectors provide relatively poor vertex determination, they are effective for peripheral collisions and give an important cross check on the vertex reconstruction obtained from the
Vertex detector and Spectrometers.

The Vertex detector was effective in determining the $z$ and $y$
coordinates of the collision vertex by connecting hits in the outer and inner silicon layers to form ``tracklets'' - two point tracks.  All possible combinations of hits on the inner and outer layer of the upper and lower Vertex detectors are constructed, producing a distribution of all possible vertex positions. When the tracklets point to the correct vertex position, a peak is observed in the vertex position distribution, whereas incorrect combinations contribute to a broad combinatorial background. The Vertex detector can only determine the $x$ coordinate with an error much larger than the beam profile because of the large dimension (24 mm and 12 mm) of the pads in that direction. Instead, this $x$ coordinate was accurately determined by using straight-line tracks formed on the basis of hits in the first 6 silicon layers of the Spectrometer arms.


The final vertex location was determined by an algorithm that performed an arbitration between the various measurements. An overall accuracy of $\sigma_x \sim$ 0.15 mm, $\sigma_y \sim$ 0.15 mm, and $\sigma_z \sim$ 0.06 mm was obtained for the 15\% most central $Au+Au$ collisions at $\sqrt{s_{_{NN}}}$=200 GeV \cite{Wozniak}. Somewhat poorer resolutions were achieved for the lower multiplicity peripheral events and at lower collision energies. The efficiency of the vertex reconstruction has been evaluated on the basis of Monte Carlo simulations using events with vertices in the range $|z|<$10 cm. In the case of the vertex detector method it was greater than 85\% for the 40-45\% centrality bin increasing to 100\% for the 30\% most central collisions at  $\sqrt{s_{_{NN}}}$=130 and 200 GeV. Because of the lower multiplicity at  $\sqrt{s_{_{NN}}}$=19.6 GeV, the vertex reconstruction efficiency was reduced in this case. 

For the {\it d+Au} and {\it p+p} collisions, the small number of tracks in the Vertex detector renders the precise Vertex detector and Spectrometer methods too inefficient for practical use. A new method, based on the energy deposition in adjacent detector pads in the Octagon detector, retains high efficiency even for low multiplicity events, but it results in a less accurate determination  of the vertex position in the range of $\sigma_z$= 0.5 to 2.0 cm depending on the event multiplicity, see Ref.~\cite{Garcia} for details. Even this approximate determination of the vertex position is sufficient for the extraction of the charged-particle multiplicity. The reconstructed multiplicity of an event is only slightly affected by the vertex position error while the more significant modifications of the shape of $dN_{ch}/d\eta$ mostly cancel by averaging over many events.

\subsection{Monte Carlo simulations}
\label{MonteCarlo}
\begin{figure}
\centerline{\epsfig{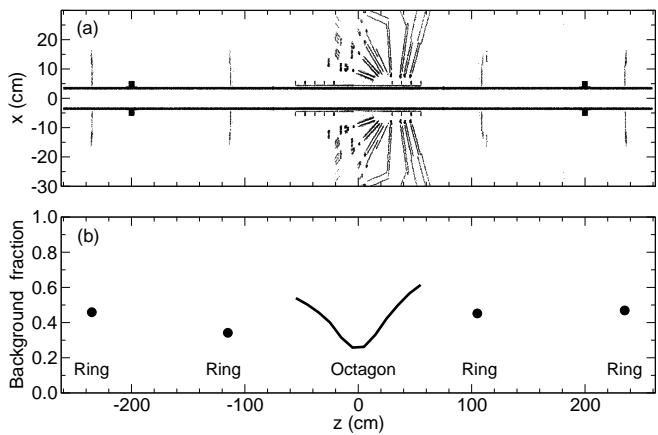}}
\caption{a) Map of simulated secondary particle sources near the horizontal plane,  $y$=0. For clarity, interactions in the air were neglected.  The beam pipe with flanges, silicon sensors and elements of their support structures are  visible. b) The ratio of hits from secondary particles to all hits in the multiplicity sensors located at different positions along the $z$-axis. The values are averaged over the azimuthal angle $\phi$. The farthest Ring counters are not shown.}
\label{Phmult_fig6}
\end{figure}

Although the reconstruction of the event multiplicity eliminates most of the background on the basis of energy deposition in the high-$\eta$ region of the Octagon detector and by accepting only tracks that point back to the vertex position in the two-layer Vertex detector, some background hits remain. This residual background is estimated on the basis of Monte Carlo simulations and appropriate corrections are then applied. Such simulations of the detector response and sources of background particles, that do not originate from the collision vertex, were carried out using the HIJING~\cite{HIJING} event generator. The GEANT 3.21~\cite{Geant} package was used to simulate the detector response and particle tracking using the detailed geometry of the PHOBOS detector. Simulations using events generated with the VENUS~\cite{Venus} and RQMD~\cite{RQMD} codes were also performed to estimate systematic errors.
All elements of the sensors and support structures were precisely modeled, 
especially those near the interaction point and along the path  
of particles to the more distant sensitive elements. In order to illustrate the degree of details of the background simulation in the PHOBOS apparatus we present in Fig.~\ref{Phmult_fig6} a map of the origin points of secondary particles leaving hits in the multiplicity detectors. One observes that even small elements, such as the Si sensors of the Spectrometer, appear in this map. Note that this figure only includes secondary particles generated near the horizontal plane. Therefore  it does not show particles created in much more massive elements, such as the magnet yokes, which are located above and below the horizontal plane.

Using Monte Carlo studies, the fraction of hits from 
secondary particles to the total number of hits in the multiplicity detectors was determined; this is shown in Fig.~\ref{Phmult_fig6}b. This fraction was smallest for hits in the Octagon sensors closest to the interaction point but it increases towards the extremities of the Octagon array because of the smaller solid angle with respect to the vertex position subtended per unit area. The ratio decreases for the Ring counters. The asymmetry with respect to $\eta$=0 was caused by secondary particles created by the magnet yokes, located in the  positive $z$ region. The values of background fraction for the two Ring detectors located at $z$=$\pm$5.05 m are identical at 0.545.

\subsection{Tracklet counting}
\label{TrackletCounting}

\begin{figure}
\centerline{\epsfig{file=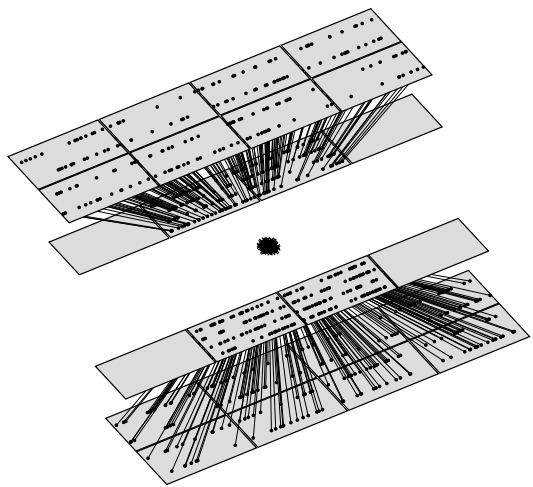,width=7cm}}
\caption{Illustration of tracklet counting using two consecutive Si layers, in this case Vertex detectors. Hits in the inner and outer Si layers are connected to form ``tracklets'' that point back to the vertex position in the center of the figure.}
\label{Phmult_fig7}
\end{figure}

Three different methods were employed to determine the charged particle multiplicity. The ``tracklet counting'' method was used in the $(\eta, \phi)$ range covered by two or more layers of silicon detectors, namely the regions covered by the Vertex detector and the Spectrometers, where it is possible to correlate ``hits'' in the two layers and thereby construct ``tracklets'' that point back to a previously determined vertex position. An illustration of   tracklets originating from the vertex position is shown in Fig.~\ref{Phmult_fig7}. The multiplicity measurements presented here were performed without the PHOBOS magnet being powered so that straight-line tracks may be assumed except for the effects of small angle scattering in the Be beam pipe and the intervening silicon layers. 

\begin{figure}
\centerline{\epsfig{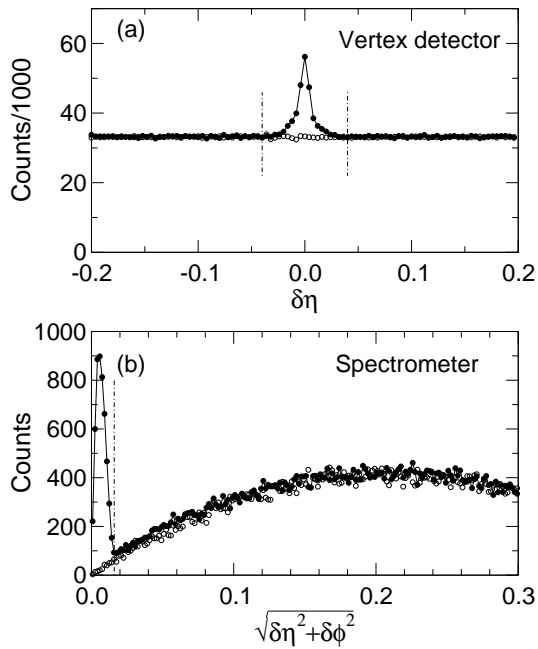}}
\caption{Illustration of the combinatorial background subtraction for non-vertex tracks for a) Vertex detector and b) Spectrometer tracklet analysis. The solid points represent the actual detector geometry, whereas the open points were obtained after a 180$^0$ rotation of one of the detector layers around the beam axis in order to estimate the combinatorial background in the acceptance region represented by vertical dashed-dotted lines.}
\label{Phmult_fig8}
\end{figure}

The Vertex detector was used in the tracklet analysis for 
particles emitted in range $|\eta|<$1. The tracklet reconstruction procedure generated ``seed'' tracks using hits in one silicon layer and the reconstructed vertex position. A seed track was extrapolated to the other silicon layer (the ``search'' layer) and the location where the seed track traversed the ``search'' layer  was compared with ``hits'' in that layer. The conditions
\begin{eqnarray}
\label{VTX_accept}
|\delta \phi|=|\phi_{\it seed}-\phi_{\it search}|<0.3\\
\nonumber |\delta \eta|=|\eta_{\it seed}-\eta_{\it search}|<0.1
\end{eqnarray}
were required for a tracklet candidate~\cite{Reuter_thesis}. This method allowed a single hit in the ``search'' layer to be shared between more than one tracklet. In such cases  only one tracklet was accepted for further analysis. The combinatorial background was estimated by        performing the same tracklet-finding procedure with the ``search'' layer rotated by $180^\circ$ in $\phi$, see Fig.~\ref{Phmult_fig8}a. After subtraction of the combinatorial background, the charged-particle multiplicity was computed taking into account the centrality and $z_{\it vtx}$ dependent detection efficiencies, which were obtained from Monte Carlo simulations.     

The tracklet analysis method was also applied to the first and second layers as well as the fifth and sixth layers of the silicon detectors in the two-arm  Spectrometer, see Fig.~\ref{Phmult_fig8}b. This allowed the multiplicity to be determined in the $0<\eta<1$ range. In this method, a slightly different tracklet acceptance criterion was used, namely
\begin{equation}
\sqrt{\delta\phi^2+\delta\eta^2} < 0.016
\end{equation} 
instead of the ``rectangular'' acceptance region described by Eq.~\ref{VTX_accept}. Tracklet multiplicities obtained from the two detectors were in excellent agreement (typically $\leq$5\%). 

Tracklets were used to obtain the centrality dependence of the charged particle multiplicity in the midrapidity region, $|\eta|<1$, and to validate measurements obtained from the hit counting and energy deposition methods described in the following section, which were applied to the full $\eta$-range subtended by single-layer silicon detectors (Octagon and Ring detectors). 

\subsection{Single Si layer analysis}

\begin{figure}
\centerline{\epsfig{file=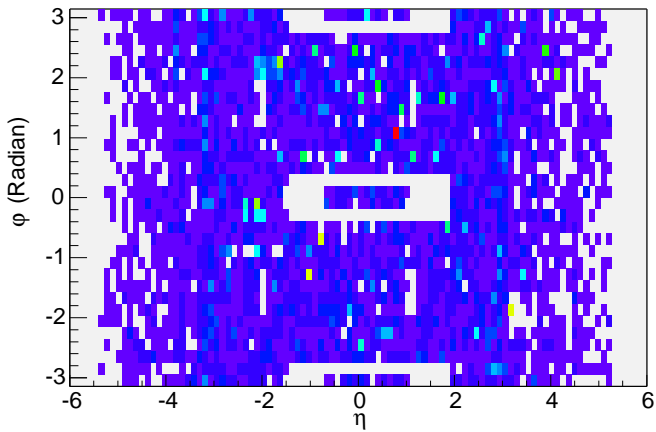,width=\columnwidth}}
\caption{(Color online) Single event energy deposition map in $\phi$ vs. $
\eta$ for a central Au+Au collision at 130 GeV. The open areas at $\phi$=0 and $\phi=\pi$ are openings in the Octagon detector in front of the Spectrometers. Small open areas at $\phi= \pm \pi/2$ and $\eta$ about -2 and 1  in front of the Vertex detectors are also visible.}
\label{Phmult_fig9}
\end{figure}

The energy deposited in the single layer of silicon detectors making up the Octagon and Ring multiplicity detectors was used to determine the charged-particle multiplicity over the
entire $\eta$ range accessible to PHOBOS.  The energy
deposited in these detectors for a typical central event is shown in Fig.\ref{Phmult_fig9}.  

The charged-particle multiplicity density was determined from the deposited energy in two independent ways. In both cases, only events with a reconstructed vertex  on the beam axis within $\pm$10 cm of the geometrical center of the multiplicity detector were used. The two methods were somewhat complementary. Hit counting was  relatively insensitive to the energy calibration since it relied only on whether the signal in a silicon pad was above a certain detection threshold. On the other hand, the energy deposition method used more of the available information at the cost of increased dependence on Monte Carlo simulations by associating the signal height in a pad with an estimated number of traversing charged particles. This method is therefore directly sensitive to the energy calibration. Both methods are described in the following sub-sections.  

\subsection{Hit counting method}
\label{HitCounting}
This method was based on counting pads with an energy deposition greater than a pre-determined minimum value. The charged-particle multiplicity was determined by summing over all pads lying in a pseudorapidity range, $\eta$, for events within a certain centrality, $b$. The pseudorapidity density $dN_{ch}/d\eta$ was obtained using the expression
\begin{equation}
\frac{dN_{ch}}{d\eta}=\frac{1}{N(b)}\sum_i^{N(b)}\left(N_{hits} \frac{O(b,\eta) \times f_{bkg}(b,\eta)}{A(z,\eta) \times \Delta \eta}\right)_i
\label{master}
\end{equation}

Here, $N(b)$ is the total number of events for a given centrality $b$, $N_{hits}$ is the number of hit pads,   $O(b,\eta)$ represents a correction due to detector occupancy,
$A(z,\eta)$ is the geometrical detector acceptance, which varies with both the
position of the collision vertex, $z$, and the pseudorapidity, and the
quantity $f_{bkg}$ takes into account the effects of backgrounds and secondary particle production. The determination of these correction terms is described below. 

\subsubsection{Occupancy}

Having corrected the deposited track energies for their angle of
incidence, and eliminated low-$\Delta E$ tracks originating from
background and secondary particles, the data must be corrected for the effects of detector occupancy.  In particular,
for central events over the entire $\eta$ range and for non-central events in the $|\eta|<3$ range there is a significant
probability that more than one particle had traversed a single
detector pad.  The detector occupancy, $O(b,\eta)$, depends upon both $b$ and $\eta$ and was determined using two different methods.

\begin{figure}
\centerline{\epsfig{file=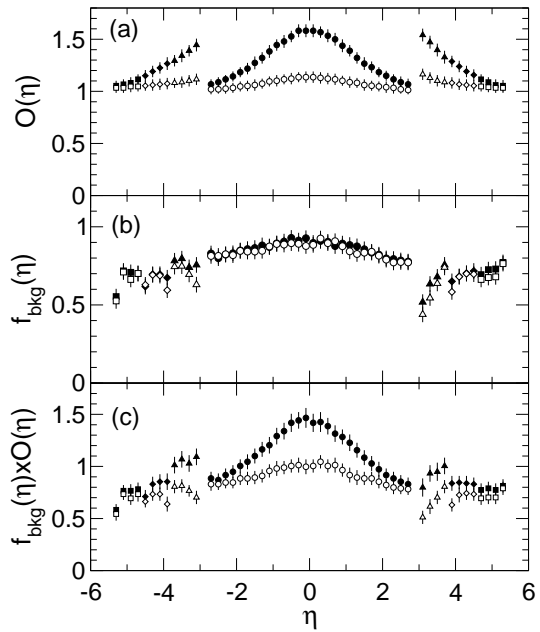,width=7cm}}
\caption{(a) The single pad occupancy correction factor $O(\eta)$ for central (0-6\%) and peripheral (45-55\%) Au+Au collisions at 130 GeV are shown as solid and open symbols, respectively. The data for the Octagon detector are shown as circles, whereas the triangles, diamonds, and squares correspond to Ring detectors at increasing distances from the vertex position. The background correction $f_{\it bkg}(\eta)$ factor and the total correction factor $O(\eta)\times f_{\it bkg}(\eta)$ are shown in panels (b) and (c), respectively, using the same symbols.}
\label{Phmult_fig10}
\end{figure}

The first method relies on the assumption that, for a given centrality, $b$, and pseudorapidity range, $\Delta \eta$, the probability that $N$ particles pass through a pad is described by a Poisson-statistical
distribution,
\begin{equation}
P(N)=\frac{\mu^Ne^{-\mu}}{N!},
\end{equation}
where $\mu$ is the average number of tracks per pad determined over an ensemble of events. The value of $\mu$ depends on both centrality and pseudorapidity, and was determined experimentally from the event sample. To determine $\mu(\eta,b)$, the number of pads, $N_{\it hit}$, in a $\Delta\eta$ range with a valid energy signal was compared to the number of pads with no hits, $N_{\it nohit}$.  The ratio $R=N_{\it hit}/N_{\it nohit}$ is related to $\mu$ by $\mu(\eta,b)=\ln[1+R(\eta,b)]$. The occupancy correction factor is then given by Poisson statistics $O(\eta,b)=\mu(\eta,b)/(\exp{\mu(\eta,b)-1)}$.
Typical values for the occupancy correction factor were between 
1.0 and 1.2 for large values of $\eta$, or for peripheral collisions,
but were larger ($O(\eta,b)\approx 1.8$) for central $Au+Au$ collisions at $\eta \approx 0$ at a collision energy of
$\sqrt{s_{_{NN}}}=200$ GeV, see Fig.~\ref{Phmult_fig10}a.

An independent experimental confirmation of the validity of the
Poisson occupancy determination was made by studies of the $\Delta E_{\it pad}$ spectra as a function of $\eta$ and $b$. After merging hits, and correcting the resulting track energies for angle of incidence, the  $\Delta E_{\it pad}$ spectra showed structures characteristic of the energy deposited by one, two, 
or more particles traversing a single pad (see Fig.\ref{Phmult_fig4}).
Each $\Delta E_{\it pad}$ spectrum was fit to a set of Landau functions convoluted with Gaussian functions to account for the intrinsic energy resolution of the pad.  From the
results of these fits, the relative contributions of $N=$1, 2, 3 hits to the $\Delta E_{\it pad}$ spectra could be determined as a function of $b$ and $\eta$.  These relative contributions then
yielded an independent occupancy correction factor $O(b,\eta)$. This method of determining the detector occupancy was used as a confirmation of the validity of the Poisson occupancy method.

\subsubsection{Background and Monte-Carlo Corrections}

The majority of particles produced by secondary or background
interactions were eliminated by requiring the value of $\Delta E_{\it pad}$ to exceed a threshold value as described above. However, there still exist additional background contributions that could not be eliminated by using measured quantities alone.  Such backgrounds included secondary particles produced in the beam pipe, in the magnet yoke, or in other detector elements in addition to those generated via feed-down from the weak decays. The background correction also accounts for the absorption of particles in the beam pipe before they had reached the silicon detectors.  To account for these effects, Monte Carlo simulations were used to determine the response of the apparatus to particles produced by a variety of event generators, as described in Sect.~\ref{MonteCarlo}. Such simulated event data, generated for a variety of collision centralities were analyzed in exactly the same manner as real data to estimate the residual background effects.  

The simulated data were first passed through the algorithm used to determine $O(b,\eta)$ following the same method as for the experimental data.  The simulated data were subsequently corrected for detector acceptance according to Eq.\ref{master}, without the term $f_{bkg}$.  The resulting $dN_{ch}/d\eta$ generally did not agree with the ``true'' known form of $dN_{ch}/d\eta$ obtained directly from the output of the event generator.  The final correction factors that were applied to the experimental data were determined from a comparison of the reconstructed Monte-Carlo $dN_{ch}/d\eta$ to the ``true'' distributions as
\begin{equation}
f_{bkg}(b,\eta) = \frac{dN_{ch}}{d\eta}({\it true})/\frac{dN_{ch}}{d\eta}({\it recon})
\end{equation}

Examples of these correction factors are shown as a function of $\eta$, for both peripheral, and central collisions, in
Fig.~\ref{Phmult_fig10}b. As expected, near $\eta=0$, $f_{\it bkg} \approx 1$ indicating that there is only a small contribution from these processes. Near $\eta$=+3 to 3.5, however, $f_{\it bkg}$  is somewhat smaller due to a large number of secondary particles produced in the steel of the magnet yoke.  As expected, the number of background and secondary
particles, as well as the number of particles absorbed in the beam pipe is proportional to the number of primary particles. This can be seen from the fact that $f_{\it bkg}$ is almost identical for peripheral and central collisions (see Fig.~\ref{Phmult_fig10}b).

\begin{figure}
\centerline{\epsfig{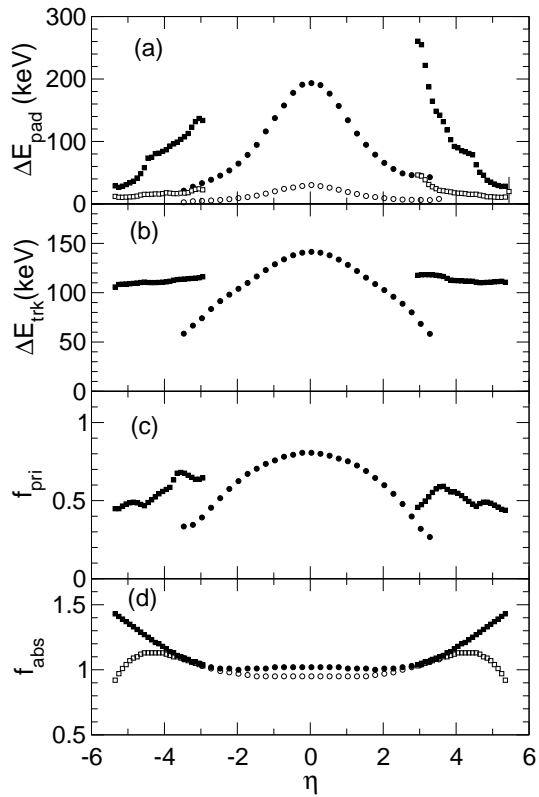}}
\caption{(a) The energy deposition per detector pad $\Delta E_{\it pad}$, (b) the energy deposition per particle track $\Delta E_{\it trk}$, (c) the fraction of primary tracks $f_{\it pri}$ , and (d) the absorption correction factor $f_{\it abs}$ are shown as a function of $\eta$ for central (0-6\%) (solid symbols) and peripheral Au+Au collisions (45-55\%) (open symbols) at 130 GeV. The circles and squares refer to simulations for the Octagon and Ring detectors, respectively }
\label{Phmult_fig11}
\end{figure}

\subsection{Energy deposition method}
\label{EnergyDeposit}

A second, largely independent, method of determining $dN_{ch}/d\eta$ used the $\Delta E_{\it pad}$ signal to estimate the number of tracks traversing a pad. This method uses the energy deposition 
in the first layer of all the silicon detectors in PHOBOS,
including not only the Octagon and Ring multiplicity detectors, but also the inner layers of the Vertex detectors, and the first Spectrometer
layers. The ``analog'' multiplicity method obtains the total
$dN_{ch}/d\eta$ from the relation
\begin{equation}
\frac{dN_{ch}}{d\eta}= \frac{1}{N(b)}\sum_i^{N(b)}\left(\sum_{pads} \frac{\Delta E_{pad} \times f_{pri} \times
f_{abs}} {\Delta 
\eta_{pad} \times \Delta E_{trk}}\right),
\end{equation}
where the sum extends over all active pads in the detectors used for the
analog analysis. $\Delta E_{pad}$ is the energy deposited in a single
pad, $\Delta E_{trk}$ is the average energy deposited in 300
$\mu$m of silicon by a single ionizing particle (SIP),
$f_{pri}(\eta)$ is the fraction of primary particle tracks
out of the total number of tracks in a given pad, $f_{abs}$ is a
correction that takes into account the effects of absorption in the
beam pipe, and $\Delta \eta_{pad}$ is the acceptance in $\eta$ for a
given pad. The quantities $\Delta
E_{trk}$ and $f_{pri}$ depend on the relative position of a given
detector sensor and the collision vertex as described below.  The
deposited energy per pad $\Delta E_{pad}$ is corrected for the angle of incidence based on the measured vertex
position. Figure \ref{Phmult_fig11} shows the pseudorapidity dependence of the various quantities in Eq. 6 for 200 GeV central (0-6\%) and peripheral (45-55\%) $Au+Au$ collisions, and that the ratio $f_{pri}/\Delta E_{trk}$ used in the calculation of $dN_{ch}/d\eta$ is nearly independent of $\eta$. 

\subsubsection{Analog method correction parameters}

The average deposited energy per track, $\Delta E_{trk}$, is
determined from Geant simulations (as described in Sect.~\ref{MonteCarlo}).  The value of $\Delta E_{trk}$ ranges between 50 and 160 keV, depending on the direction of the track traversing the sensor and the mixture of primary and secondary particles. Note that the average energy deposition, $\Delta E_{trk}$, is substantially larger than the most probable value due to the asymmetric nature of the Landau distribution as illustrated in Fig.~\ref{Phmult_fig4}. The $\eta$ dependence in the Octagon primarily reflects a sensitivity to the broadening of the energy-loss distribution as the particles encounter more silicon with increasingly shallow angles of incidence.  The additional $\eta$-dependence is due to the changing mix of particle types, particle momenta, and the number of secondary particles produced within the volume of the silicon. These variations do not depend
strongly on the event generator used to obtain the primary particle
distributions. The values obtained here are to be compared with the mean energy deposition
value of 117 keV noted in the Particle Data Book\cite{PDB} for minimum
ionizing particles normally incident on a 300 $\mu$m thick wafer of
silicon.  The value of $\Delta E_{trk}$ is calculated for each sensor
in the first silicon layer in the PHOBOS detector, and its dependence on $\eta$ is fitted with a polynomial parameterization for computational purposes.

The quantity, $\Delta E_{trk}$, was also measured using the actual data under low occupancy conditions, corrected for multiple occupancy. The final validation of the analog method parameters, and the systematic uncertainty assignments, are provided by observing the close agreement between the two (and three, where available) analysis methods.

The fraction of primary particles to the total number of particles
$f_{pri}=\frac{N(primary)}{N(total)}$ is determined  from
simulations using events produced by HIJING (see Sect.~\ref{MonteCarlo}), and includes secondary particles produced by interactions with material as well as
feed-down from weak decays.  The values of this parameter vary from
approximately 0.9 near $\eta$=0, to $\approx 0.3$ near the
ends of the Octagon, or for the Ring detectors closest to the vertex
position.  The dependence of $f_{pri}$ on pseudorapidity, centrality and vertex position is similar to that of the background correction factors from the hit counting analysis described above.  This quantity is also
parameterized for computational purposes.

The absorption coefficient $f_{abs}$ takes into account the absorption of particles in the material traversed before encountering the silicon detectors, chiefly the 1 mm Be beam pipe. As such, this correction has a $\cosh\eta$ dependence, with the value of $f_{abs}$ being approximately 0.98 near $\eta$=0. The acceptance of each pad, $\Delta \eta$, is calculated as a function of $\eta$ from the measured PHOBOS geometry.   

\begin{figure}[hbt]
\centerline{\epsfig{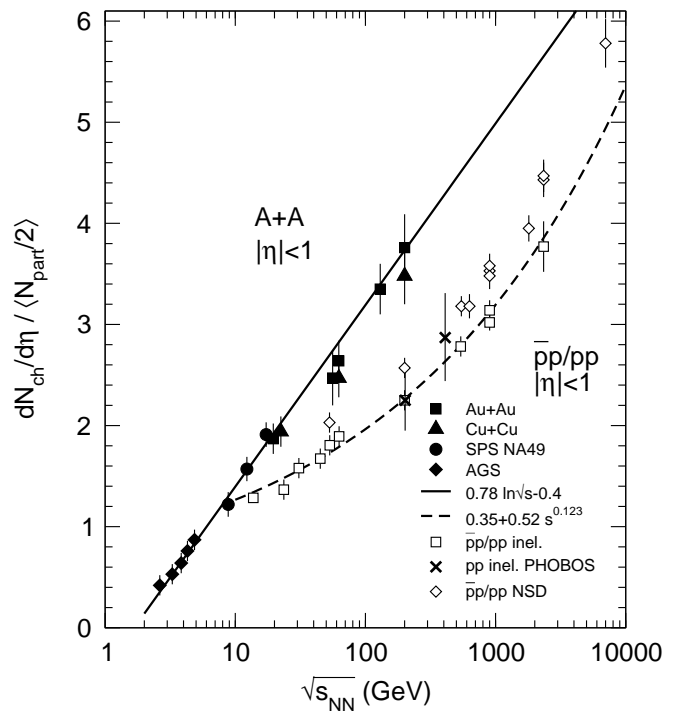}}
\caption{The energy dependence of the participant-scaled charged particle multiplicity 
$dN_{ch}/d\eta/\langle N_{part}/2 \rangle$ at midrapidity 
$|\eta|<1$ is shown for 0-6\% central nucleus-nucleus collisions (solid symbols) and compared to $\bar p p/pp$ non-single diffractive (NSD) (open diamonds) \cite{PRD41_2330,Alner,NPB129_365,ALICE} and inelastic (open squares) \cite{Morse,NPB129_365,Alner,ALICE,CMS,CMS_7TeV}) and (crosses) present work. The solid squares and triangles represent results from the PHOBOS experiment for $Au+Au$ and $Cu+Cu$ collisions, respectively. The solid circles and solid diamonds are obtained from SPS $Pb+Pb$  \cite{NA49} and AGS $Au+Au$ \cite{AGS}   collisions, respectively. The 56 GeV $Au+Au$ data point is from Ref.~\protect\cite{PH1}. The solid line is a linear fit to the $Au+Au$ and $Pb+Pb$ data, whereas the dashed curve is a fit to the inelastic $\bar pp/pp$ data.
See text for details. 
}
\label{Phmult_fig12}
\end{figure}

\section{Results}
\label{Results}

The results are for all charged particles excluding weak decays and corrected for missing low $p_T$ particles. Although the tracklet and single-layer measurements of the charged particle multiplicity are in good agreement, we present the results separately, because of their different range of applicability. For the centrality and energy dependence of the charged particle density at midrapidity the tracklet method is more accurate, whereas the overall pseudorapidity distribution can be obtained only from the hit-counting and energy deposition analysis which can be applied away from $\eta=0$.

\begin{figure}[hbt]
\centerline{\epsfig{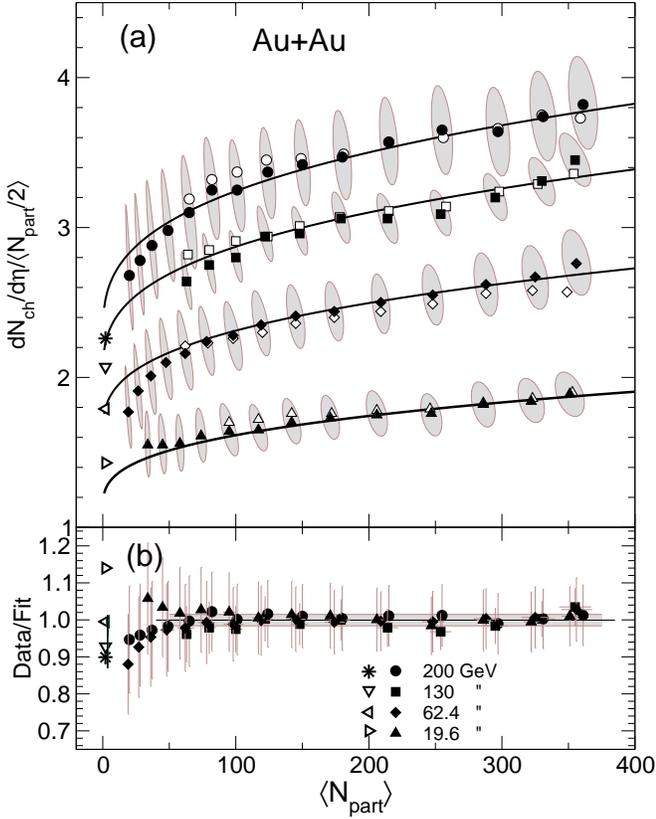}}
\caption{
Panel a: The charged particle multiplicity at midrapidity $|\eta|<1$ per participant pair, $dN_{\it ch}/d\eta/\langle N_{\it part}/2 \rangle$ obtained from the ``tracklet'' analysis is shown as a function of $\langle N_{\it part}\rangle$  (solid symbols) for $Au+Au$ collisions. The shaded ovals indicate the 90\% confidence limit systematic errors. For comparison, open points show the ``single Si layer'' analysis.
The solid curves represent fits to the data using the form given in Eq.~\ref{factorization} - \ref{f_and_g} (excluding the $pp$ points). The $pp$ points at $\langle N_{part}\rangle$=2 were interpolated using the fit to the inelastic $pp/\overline pp$ data, see Fig.~\ref{Phmult_fig12}. See text for a discussion of panel (b).
}
\label{Phmult_fig13}
\end{figure}

\begin{figure}
\centerline{\epsfig{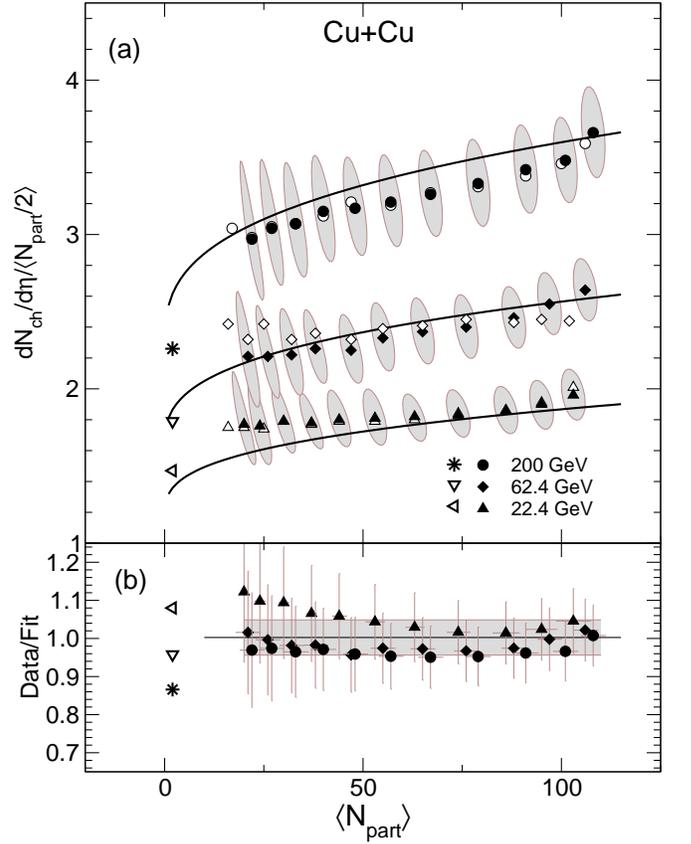}}
\caption{Same as Fig.~\ref{Phmult_fig13}, but for $Cu+Cu$ collisions.}
\label{Phmult_fig14}
\end{figure}

\subsection{Charged particle production at midrapidity.}


\subsubsection{Energy dependence}
\label{EnergyDependence}

The energy dependence of the charged particle multiplicity, normalized by the number of participant pairs, is shown in Fig.~\ref{Phmult_fig12}a for central (0-6\%) $Au+Au$ (solid squares and diamonds), $Cu+Cu$ (solid triangles), $Pb+Pb$ (solid circles)~\cite{NA49} collisions. The $Au+Au$ and $Cu+Cu$ data shown in this figure and listed in Table~\ref{table0} represent an average of the results from the tracklet and single Si layer analyzes. The heavy-ion data are compared to the $dN_{\it ch}/d\eta$ data obtained in non-single diffractive (open diamonds)~\cite{NPB129_365,Alner,PRD41_2330}, and inelastic (open squares~\cite{Morse,NPB129_365,Alner}) $\bar p+p$ or $p+p$-collisions. One observes a strong enhancement of the midrapidity charged particle production per participant pair in $Au+Au$ collisions compared to $\bar p+p$ of up to 70\%. 

The overall energy dependence of $\frac{2}{\langle N_{\it part}\rangle}\frac{dN_{\it ch}}{d\eta}$ is logarithmic for A+A collisions. The solid line, given by
\begin{equation}
\frac{2}{\langle N_{\it part}\rangle}\frac{dN_{\it ch}}{d\eta} = 0.78 \ln(\sqrt{s})-0.4,
\label{eq7}
\end{equation}
is seen to describe the $Au+Au$ and $Pb+Pb$ data quite accurately over the two orders of magnitude of collision energy, but it also appears that the $Cu+Cu$ data fall slightly below this trend. Also the 56 and 62.4 GeV $Au+Au$ points fall slightly below this line, which may indicate a curvature to the collision energy dependence. The dashed curve is a fit to the inelastic $\bar pp$ and $pp$ data, namely $dN_{ch}/d\eta /\langle N_{part}/2 \rangle = 0.35+0.52 s^{0.123}$.

\begin{table}  
\caption{Summary of the midrapidity $\frac{dN_{\it ch}}{d\eta}|_{|\eta|<1}/\langle N_{part}/2\rangle$ charged particle multiplicity for $Au+Au$ and $Cu+Cu$ for 0-6\% central collisions. The data are averaged over those obtained from the tracklet counting and single Si layer analysis. Errors represent averages of the 90\% C.L. systematic errors for the two methods. Statistical errors are negligible.}
\vspace{3mm}
\begin{tabular}{cccc}
\tableline
System 	&$\sqrt{s_{_{NN}}}$(GeV) & $\langle N_{\it part}\rangle$&$\frac{dN_{\it ch}/d\eta|_{|\eta|<1}}{\langle N_{\it part}/2\rangle}$\\

\hline
Au+Au	&	19.6	&	337	&	1.87$\pm$0.15\\
"	&	56	&	330	&	2.47$\pm$0.27\\
"	&	62.4	&	338	&	2.64$\pm$0.20\\
"	&	130	&	342	&	3.35$\pm$0.25\\
"	&	200	&	345	&	3.76$\pm$0.33\\
\hline
Cu+Cu	&	22.4	&	99	&	1.94$\pm$0.15\\
"	&	62.4	&	96	&	2.47$\pm$0.19\\
"	&	200	&	100	&	3.48$\pm$0.28\\
\tableline
\end{tabular}
\label{table0}
\end{table}

\subsubsection{Centrality dependence and factorization}

The midrapidity charged-particle multiplicities normalized to the number of participant pairs, $\langle N_{\it part} \rangle/2$, are shown for $Au+Au$ collisions (solid symbols) in Fig.~\ref{Phmult_fig13} as a function of centrality of the collision expressed by $\langle N_{\it part} \rangle$ and listed in Table~\ref{table1}, column 4 (see Appendix). The corresponding data for $Cu+Cu$ collisions are given in Fig.~\ref{Phmult_fig14}a and Table~\ref{table2} (see Appendix). The shaded ovals represent estimates of the 
90\% confidence level systematic errors
in the measured $dN_{ch}/d\eta$ values and the calculated number of participant pairs, $\langle N_{\it part}\rangle/2$. One observes that the normalized charged particle production at all energies increases with $\langle N_{\it part} \rangle$ and exceeds the measurements for $pp$ and $\bar pp$ inelastic collisions represented by stars and open triangles at $N_{part}$=2 in Figs.~\ref{Phmult_fig13} and~\ref{Phmult_fig14}. The 200 GeV $\bar pp$ value of 2.25$\pm$0.1 was measured in Ref~\cite{Alner}, whereas in 62.4 GeV collisions a value of 1.89$\pm$0.1 was measured in Ref.~\cite{NPB129_365}. A fit to the multiplicities observed for inelastic $\bar pp$ collisions Ref. \cite{Morse,NPB129_365,Alner} was obtained, see Fig.~\ref{Phmult_fig12}, and used to derive data points for 19.6, 22.4, and 130 GeV collisions. 

As demonstrated in a previous PHOBOS  publication~\cite{PH17}, the collision energy and centrality dependences of charged particle production in $Au+Au$ collision at midrapidity exhibit factorization such that

\begin{equation}
\frac{2}{\langle N_{\it part}\rangle}\frac{dN_{\it ch}|_{|\eta|<1}}{d\eta} =f(s) \times g(N_{\it part}).
\label{factorization}
\end{equation}

For $Au+Au$ collisions we find that the data are very well described by the functions
\begin{eqnarray}
f(s) = 0.0147 (\ln s)^2 +0.6 \\
 g(N_{\it part})=1+0.095 N_{\it part}^{1/3}
\label{f_and_g}
\end{eqnarray}   
shown as solid curves in Fig.~\ref{Phmult_fig13}a.   

The ratio of the data to this fit is shown in Fig.~\ref{Phmult_fig13}b. The small standard deviation, $\sigma=0.0155$ (shaded band) from the mean value $\langle data/fit \rangle$=0.9993 of all $Au+Au$ data points (horizontal line) illustrates the accuracy of the factorization; $\sigma$ is much smaller than the estimated total error (systematic and statistical) on the data points. Note that the functional form of the energy dependence chosen here is different from the overall trend discussed in the previous sub-section since it provides a slightly better fit over the limited range of collision energy for the $Au+Au$ data shown in Fig.~\ref{Phmult_fig13}; this choice illustrates better the high degree of energy-centrality factorization observed in these data.
 
For $Cu+Cu$ collisions the same energy dependence function $f(s)$ applies, whereas the $N_{part}$ dependence is given by

\begin{equation}
g(N_{\it part})=1+0.129 N_{\it part}^{1/3},
\end{equation}
as shown by solid curves in Fig.~\ref{Phmult_fig14}a. Again, Fig.~\ref{Phmult_fig14}b displays the ratio of data to this fit, which exhibits only a small deviation from unity. The line represents the average value (R=1.0026)of this ratio over all data points and the shaded band the corresponding standard deviation of $\sigma$=0.047.

\begin{figure}
\centerline{\epsfig{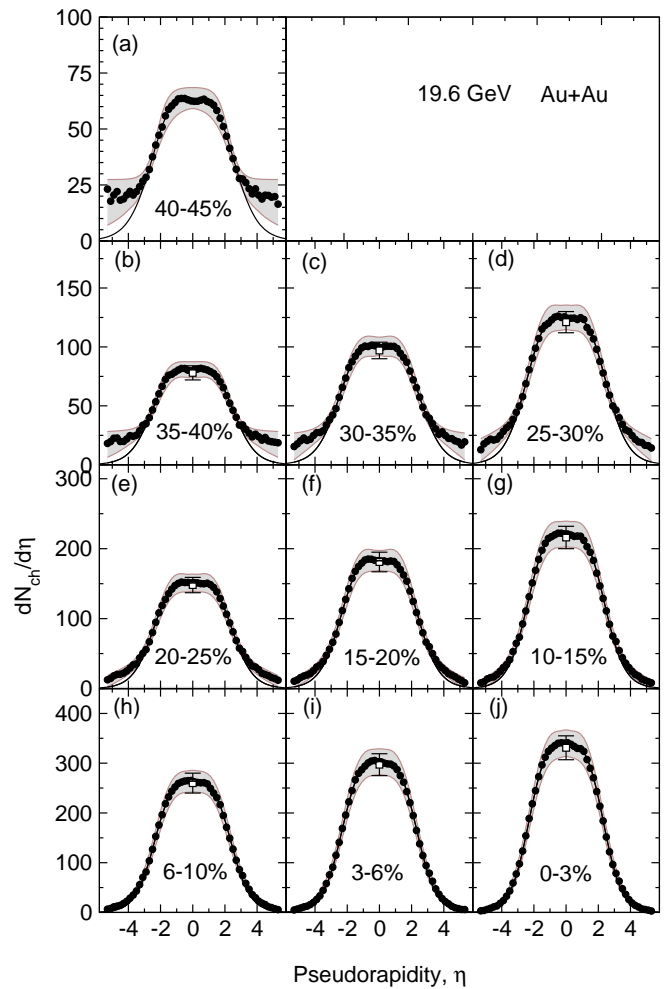}}
\caption{$dN_{\it ch}/d\eta$ vs $\eta$ (solid points) for ten centrality bins representing 45\% of the total cross section for $Au+Au$ collisions at $\sqrt{s_{_{NN}}}$=19.6 GeV. The solid curve is a fit  to the data within the $-3 <\eta < 3$ region using Eq.~\ref{WS_fit} (see text for details). The shaded band represent 90\% C.L. systematic errors. The open points were obtained by the tracklet analysis in the range $|\eta|<1$. }
\label{Phmult_fig16}
\end{figure}

\subsection{Pseudorapidity distributions}

\begin{figure}
\centerline{\epsfig{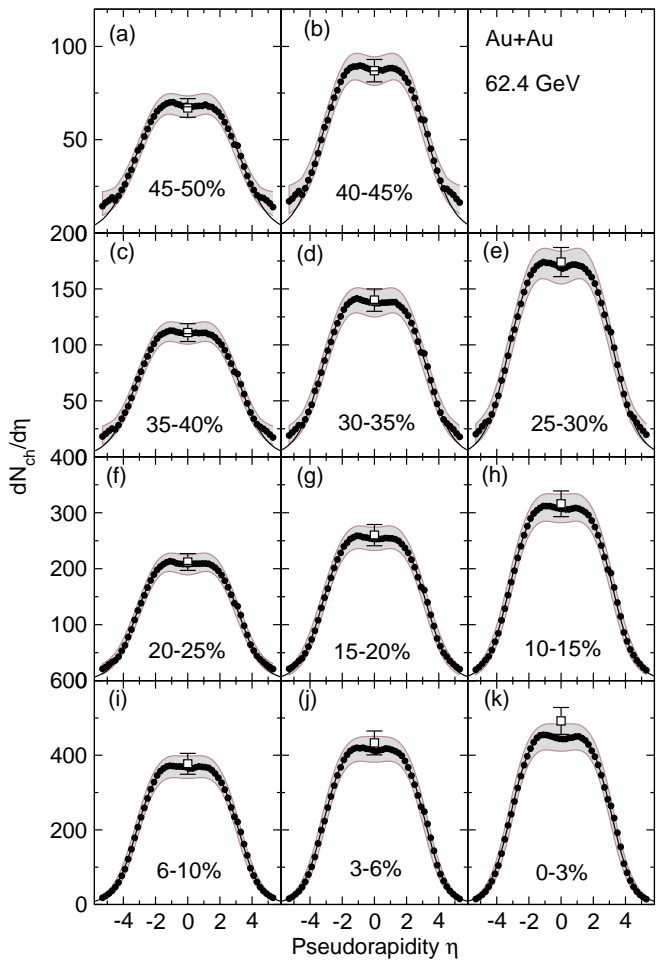}}
\caption{$dN_{\it ch}/d\eta$ vs $\eta$ (solid points) for eleven centrality bins representing 50\% of the total cross section for $Au+Au$ collisions at $\sqrt{s_{_{NN}}}$=62.4 GeV. The solid curve is a fit to the data within the $-4.2 <\eta < 4.2$ region using Eq.~\ref{WS_fit} (see text for details). The shaded band represent 90\% C.L. systematic errors. The open points were obtained by the tracklet analysis in the range $|\eta|<1$.}
\label{Phmult_fig17}
\end{figure}

The final pseudorapidity distributions, $dN_{\it ch}/d\eta$, are obtained by a simple equal weight averaging of the results from the hit-counting and energy deposition methods described in 
Sect.~\ref{HitCounting} and Sect.~\ref{EnergyDeposit}, respectively. The distributions are shown as solid points in Figs.~\ref{Phmult_fig16}-\ref{Phmult_fig19} for {\it Au+Au} collisions and in Figs.~\ref{Phmult_fig20}-\ref{Phmult_fig22} for {\it Cu+Cu} collisions for different centrality bins. The shaded bands represent the range of systematic errors to the 90\% confidence-level.

All  $dN_{\it ch}/d\eta$ distributions exhibit a plateau around $\eta\sim 0$, the range of which increases with collision energy followed by a smooth fall-off to higher values of $|\eta|$. The fall-off, which is associated with the extended longitudinal scaling region (see Sect.~\ref{els} and Ref.~\cite{PH23}), is increasing in range with energy. It is also apparent that the level of the central plateau increases with both centrality and collision energy. Note that an earlier analysis of the 19.6 GeV {\it Au+Au} data~\cite{PH10} gave up to 10\% lower values. A re-analysis of this data set using an improved centrality determination and an improved dead-channel map led to the more reliable measurement presented here.

\subsubsection{Total charged particle multiplicity}

Nearly all the charged particles fall within the acceptance of the PHOBOS multiplicity array.  Therefore, it is appropriate to integrate the $dN_{\it ch}/d\eta$ distributions in order to estimate the total number of charged particles emitted in the collision, which was done using three different methods as illustrated in Fig.\ref{Phmult_fig15}. The first estimate, $N_{ch}|_{|\eta|<5.4}$, is a simple integration of $dN_{\it ch}/d\eta$ over the $\eta$ acceptance, $-5.4 < \eta < 5.4$ of the multiplicity array corresponding to the shaded areas in Fig.~\ref{Phmult_fig15}. The results of this analysis are listed in column 4 of Tables~\ref{table3} and \ref{table4} (see Appendix) and plotted as open circles in Figs.~\ref{Phmult_fig23} and \ref{Phmult_fig24}.  

\begin{figure}[tb]
\centerline{\epsfig{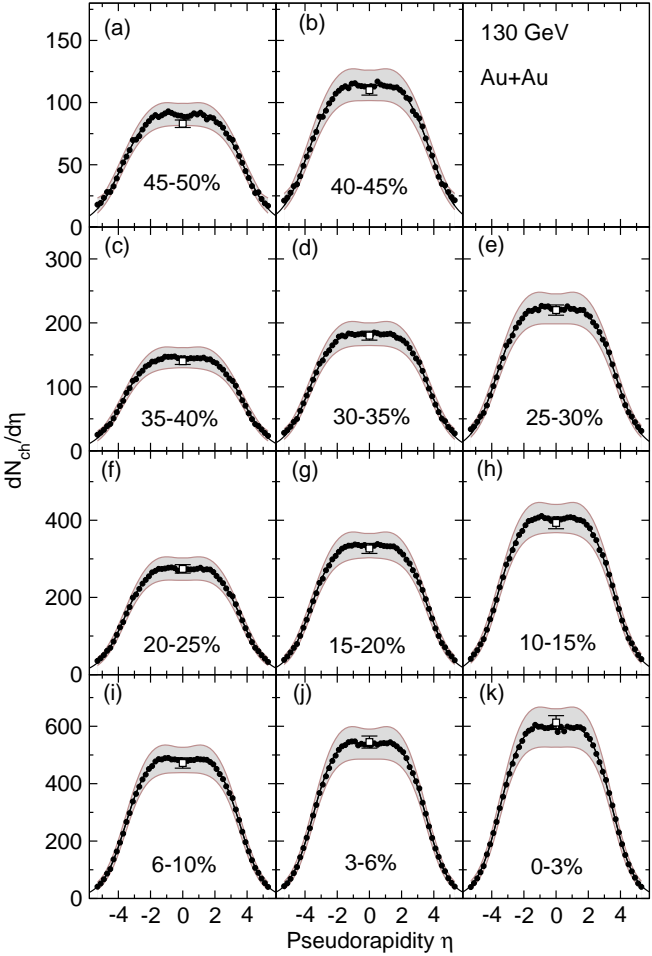}}
\caption{Same as Fig.~\ref{Phmult_fig17} but for $Au+Au$ collisions at $\sqrt{s_{_{NN}}}$ = 130 GeV.  The solid curves represent best fits to the data over the region -4.9$<\eta<$4.9 using Eq.~\ref{WS_fit}  (see text) and the shaded regions represent the systematic error band at 90\% confidence limit. The open points were obtained by the tracklet analysis in the range $|\eta|<1$.}
\label{Phmult_fig18}
\end{figure}

\begin{figure}
\centerline{\epsfig{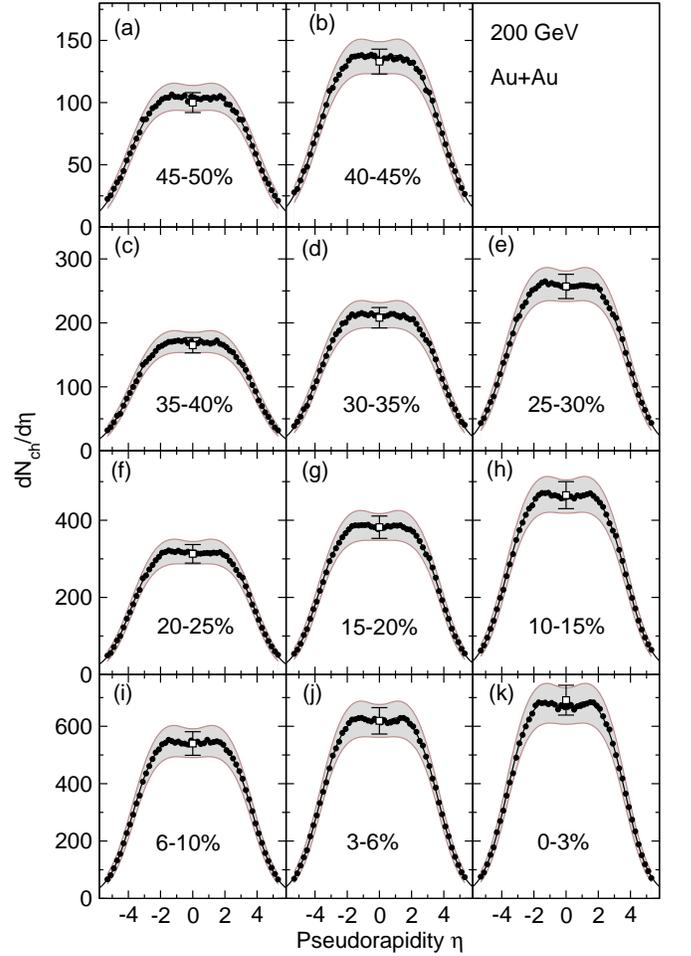}}
\caption{Same as Fig.~\ref{Phmult_fig18} but for $Au+Au$ collisions at $\sqrt{s_{_{NN}}}$ = 200 GeV. The solid curves represent best fits to the data over the full $\eta$ range using Eq.~\ref{WS_fit} and the shaded regions represent 90\% C.L. systematic errors. The open points were obtained by the tracklet analysis in the range $|\eta|<1$.}
\label{Phmult_fig19}
\end{figure}

\begin{figure}

\centerline{\epsfig{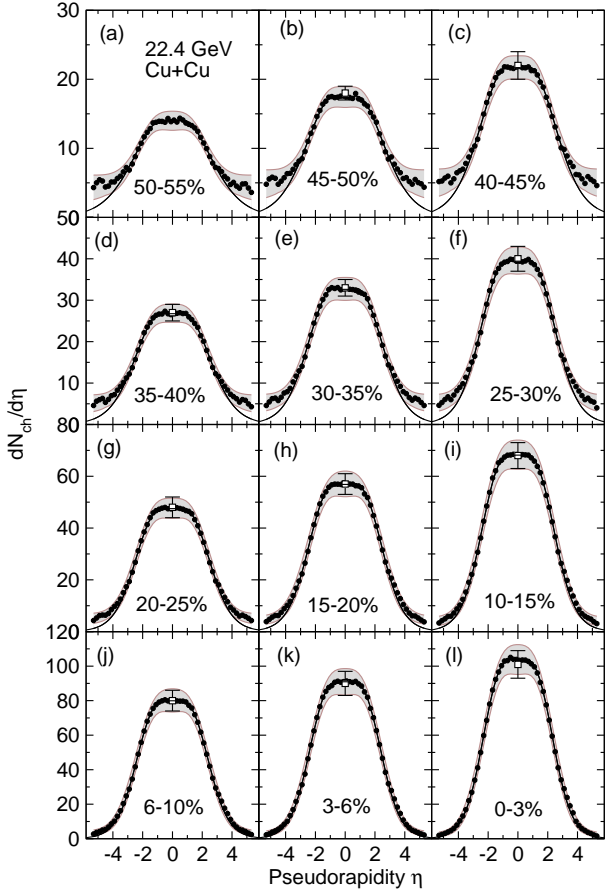}}
\caption{$dN_{\it ch}/d\eta$ vs $\eta$ (solid points) for twelve centrality bins representing 55\% of the total cross section for $Cu+Cu$ collisions at $\sqrt{s_{_{NN}}}$=22.4 GeV. The solid curve is a symmetric double quasi-Gaussian function fit to the data within the $-3.2 <\eta < 3.2$ region. The shaded band represent 90\% C.L. systematic errors. The open points were obtained by the tracklet analysis in the range $|\eta|<1$.}
\label{Phmult_fig20}
\end{figure}

By inspection of Figs.~\ref{Phmult_fig16} and \ref{Phmult_fig20} showing the $dN_{ch}/d\eta$ distributions for {\it Au+Au} and {\it Cu+Cu} collisions at $\sqrt{s_{_{NN}}}$=19.6 and 22.4 GeV, respectively, it is apparent, however, that large high-$\eta$ tails develop as one moves toward more peripheral collisions. One possibility is that these tails represent charged particles emitted from collision spectators, which travel in a very forward direction after being sheared off from the participant part of the incoming nuclei during the collision. The fact that the tails become more prominent for peripheral collisions and lower collision energies is consistent with this picture. Under this assumption, we estimate the multiplicity of charged particles originating from the collision zone by excluding these tails using the following procedure. We have found that for central collisions of {\it Au+Au}  at 62.4, 130 and 200 GeV and {\it Cu+Cu} at 62.4 and 200 GeV, the $dN_{\it ch}/d\eta$ distributions are well reproduced by

\begin{equation}
\frac{dN_{ch}}{d\eta} = \frac{c\sqrt{1-1/(\alpha \cosh \eta)^2}}{1+e^{(|\eta|-\beta)/a}},
\label{WS_fit}
\end{equation}
where c, $\alpha$, $\beta$, and $a$ are fit parameters. The quality of such fits, shown as solid  curves, can be examined in {\it e.g.} Figs.~\ref{Phmult_fig18}k and \ref{Phmult_fig19}k.

\begin{figure}
\centerline{\epsfig{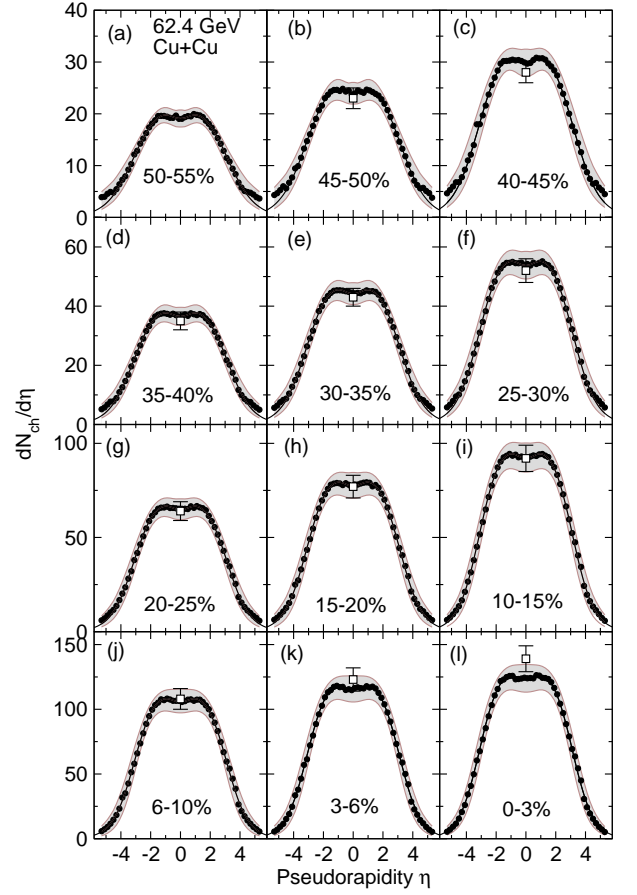}}
\caption{$dN_{\it ch}/d\eta$ vs $\eta$ (solid points) for twelve centrality bins representing 55\% of the total cross section for $Cu+Cu$ collisions at $\sqrt{s_{_{NN}}}$=62.4 GeV. The solid curve is a fit to the data within the $-4.2 <\eta < 4.2$ region using Eq.~\ref{WS_fit}. The shaded band represent 90\% C.L. systematic errors. The open points were obtained by the tracklet analysis in the range $|\eta|<1$.}
\label{Phmult_fig21}
\end{figure}

\begin{figure}
\centerline{\epsfig{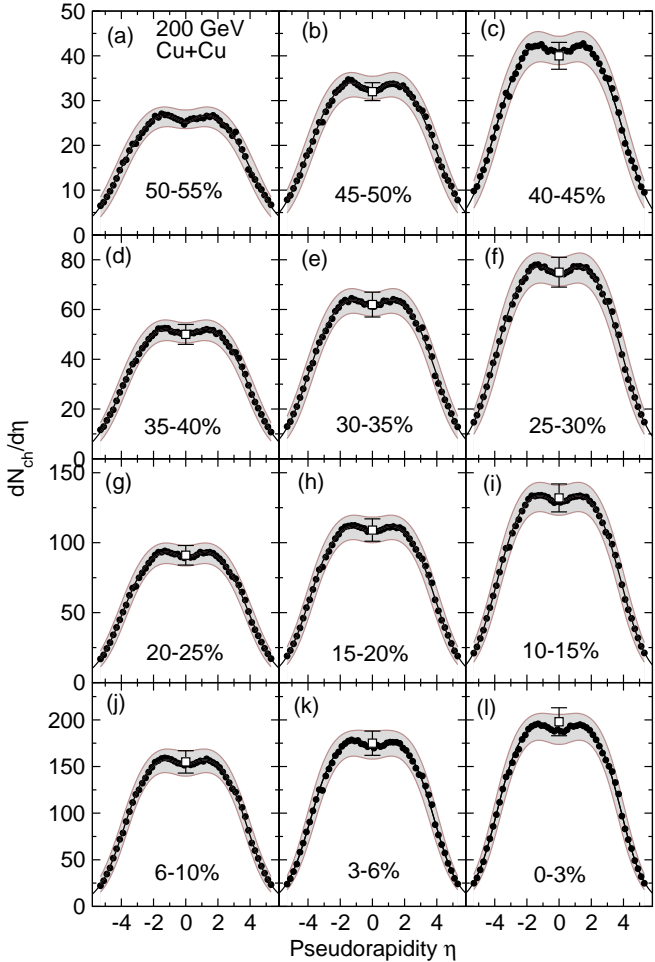}}
\caption{Same as Fig.~\ref{Phmult_fig21} but for $\sqrt{s_{_{NN}}}$ = 200 GeV. The solid curve is a fit to the data over the full $\eta$ region using Eq.~\ref{WS_fit}. The open points were obtained by the tracklet analysis for events in the range $|\eta|<1$.}
\label{Phmult_fig22}
\end{figure}

\begin{figure}
\centerline{\epsfig{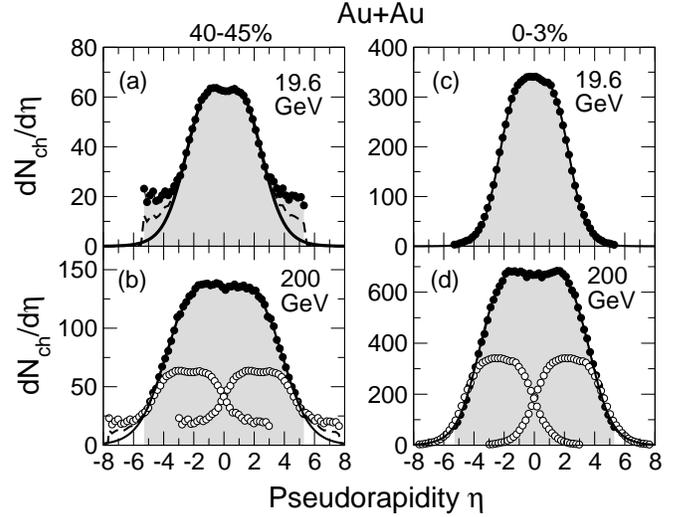}}
\caption{Illustration of the methods used for estimating the total number of charged particles from the measured $dN_{\it ch}/d\eta$ vs $\eta$ (solid points) distributions. The solid curve represents a fit to the data within the region $-y_{beam} <\eta < y_{beam}$ using the functional form of  Eq.~\ref{WS_fit}. The open circles and the shaded regions are explained in the text.}
\label{Phmult_fig15}
\end{figure}

It has been demonstrated~\cite{PH10,PH34} that {\it Au+Au} and {\it Cu+Cu} collisions rather accurately obey extended longitudinal scaling ({\it a.k.a.} limiting fragmentation scaling), even for non-central collisions, such that the multiplicity in the range around $\eta=y_{\it beam}$ is independent of collision energy. Consequently, we expect that the particle production in the range $-y_{\it beam} < \eta < y_{beam}$ is strongly dominated by participant collisions ({\it i.e.} excluding spectator emission). A fit to the data within this pseudorapidity range at all collision energies was therefore assumed to describe the contribution to the total charged particle multiplicity from participant collisions also outside this interval. The solid black curves in Figs.~\ref{Phmult_fig16} - \ref{Phmult_fig22} represent these fits to the data using the functional form and fit interval specified in the figure captions. The estimated total charged particle multiplicities for participant collisions obtained from this procedure are denoted $N_{ch}^p$ and are listed in Tables~\ref{table3} and \ref{table4} (see Appendix) and shown as solid circles in Figs.~\ref{Phmult_fig23} and  \ref{Phmult_fig24}. The quantity $N_{ch}^p$ should represent the most reliable estimate of the total number of charged particles emitted from the overlap zone between the colliding nuclei.

The third estimate, $N^{ch}_{tot}$ represents an extrapolation outside the measured $\eta$ region,  which does not exclude contributions from spectator emission. This method was used in all earlier PHOBOS publications, {\it e.g.} ~\cite{PH10,PH24,PH25,PH29,PH34}.
For the lowest collision energy, 19.6 GeV and 22.4 GeV for $Au+Au$ and $Cu+Cu$ collisions, respectively, an average of $N_{ch}|_{|\eta|<5.4}$ and $N_{ch}^p$ is used. This corresponds to the area underneath the dashed curve in Fig.~\ref{Phmult_fig15}a. For higher collision energies, the experimental data (solid points in Fig.~\ref{Phmult_fig15}b,d) are extended outside the measured $\eta$ region by shifting the low energy distributions by $\Delta \eta = \pm [y_{beam}-y_0]$, where $y_{beam}$ is the beam rapidity at the collision energy in question and $y_0$ is the beam rapidity at the lowest collision energy. This extension of the data is thus based on the assumption that $dN_{ch}/d\eta$ for $\eta>y_{beam}$ is independent of collision energy - the limiting fragmentation hypothesis. The $N_{ch}^{tot}$ estimate is thus the integral of the average of this extended distribution and the fit using Eq.~\ref{WS_fit}(dashed curve in Fig.~\ref{Phmult_fig15}b).

Tables~\ref{table3} and \ref{table4} (see Appendix) summarize the total charged particle multiplicity results for {\it Au+Au} and {\it Cu+Cu} collisions, respectively. The estimated average number of participants associated with each centrality bin was obtained from Glauber model (Monte Carlo version) \cite{Glauber} and listed in column two. Column three lists the full width at half maximum (FWHM) of the $dN_{\it ch}/d\eta$ distributions, whereas the three total multiplicity estimates discussed above are listed in columns 4-6.


The upper panels of Figs.~\ref{Phmult_fig23} and ~\ref{Phmult_fig24} display the values $N_{ch}^p$ (solid points) and $N_{\it ch}|_{|\eta|<5.4}$ (open circles). In all cases one observes participant scaling (Ref.~\cite{Busza}), an essentially linear dependence on $\langle N_{part} \rangle$. This is illustrated more clearly in the middle panels, where the participant-scaled results, $dN_{ch}/d\eta/\langle N_{part}/2 \rangle$ are seen to be essentially independent of $\langle N_{part}/2 \rangle$ and exceeding the values obtained in $pp/\overline pp$ collisions. We observe that this quantity is almost constant with collision centrality.  
It is interesting to note that the normalized particle production in heavy-ion collisions is larger by about 40\% than those of $\bar pp$ collisions (solid squares)\cite{Alner} and $pp$ collisions (solid diamonds)~\cite{NPB129_365}. 

The widths of the $dN_{\it ch}/d\eta$-distributions, represented by the Full Width at Half Maximum (FWHM) are shown in the bottom panels in Figs.~\ref{Phmult_fig23} and ~\ref{Phmult_fig24} for $Au+Au$ and $Cu+Cu$ collisions, respectively. The FWHM exhibit a decline with centrality, which indicates that the increased particle production with centrality preferentially occurs in the midrapidity region. Note also that the FWHM  for $\bar pp$ reactions  at 200 GeV  and $pp$ reactions at 62.4 GeV and 19.6 GeV  follow the trend of the $Au+Au$ data extrapolated to $\langle N_{\it part}\rangle=2$. A similar trend is found for $Cu+Cu$ collisions.

\subsubsection{Energy dependence}

In this sub-section we describe a simple semi-empirical expression for the total charged particle multiplicity in central heavy-ion collisions. It is motivated by the observation that these distributions are largely characterized by a midrapidity plateau, the height of which is well described by the empirical relation Eq.~\ref{eq7} and the fact that the width increases with collision energy such that the extended longitudinal scaling \cite{PH10} is fulfilled. A linear fit to the $Au+Au$ data at 0-3\% centrality in the region $2.0<\eta<5.0$ shows that it is well reproduced by the relation $dN_{ch}/d\eta = \alpha(y_{beam}+\eta_0-\eta)$, and $dN_{ch}/d\eta = \alpha(y_{beam}+\eta_0+\eta)$ for $-5.0<\eta<-2.0$, where $\eta_0$=0.11 represents a small pseudorapidity offset and the slope in the fragmentation region is $\alpha$=205. Since the mid-rapidity region is approximately flat, it is appropriate to use a trapezoidal shape of the $dN/d\eta$ distributions which leads to an expression of the form
\begin{equation}
N_{\it ch}^{tpz} = \frac{dN_{\it ch}|_0}{d\eta}(2\eta_0+2y_{\it beam}-\frac{\langle N_{part}\rangle}{2\alpha}\frac{dN_{\it ch}|_0}{d\eta}),
\end{equation}   
where the term $2y_{\it beam}$ accounts for the increased width of the distribution as a function of collision energy, see Ref.~\cite{Bari} for a similar analysis of the 0-6\% centrality bin. Using the approximation $y_{beam} \simeq \frac{1}{2}\ln s_{_{NN}} -\ln(m_0c^2)$, which is valid for $\sqrt{s_{_{NN}}}\gg m_0$ ($m_0$ being the nucleon mass) one obtains

\begin{equation}
\frac{N_{ch}^{tpz}}{\langle N_{part}/2 \rangle} \simeq   0.26 (\ln s_{_{NN}})^2+0.01 \ln s_{_{NN}}-0.28,
\label{trap}
\end{equation}
which is compared to experimental data for $Au+Au, Pb+Pb$, and $Cu+Cu$ collisions in Fig.~\ref{Phmult_fig25} (dashed curves). This expression reproduces the PHOBOS data quite well, as expected, but underestimates the charged particle production at lower energies, presumably due to a breakdown of the trapeziodal shape approximation at these lower energies. This derivation does, however, predict the leading $(\ln s_{_{NN}})^2$ term. We find that the expression 

\begin{equation}
\frac{N_{ch}^{tpz}}{\langle N_{part}/2 \rangle} =   0.26 (\ln s_{_{NN}})^2 +0.12,
\label{Nch_expression}
\end{equation}
gives an excellent description of the overall charged particle production in heavy-ion collisions over the full energy range from $\sqrt{s_{_{NN}}}$ = 2.4 GeV to  $\sqrt{s_{_{NN}}}$=200 GeV (solid lines in Fig.~\ref{Phmult_fig25}).

\begin{figure}
\centerline{\epsfig{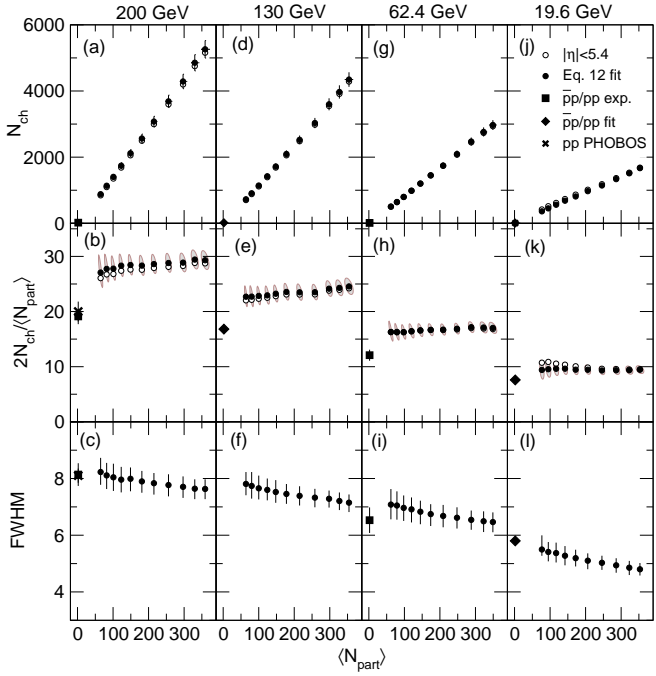}}
\caption{The total charged particle multiplicity, $N_{ch}$, the total charged particle multiplicity per participant pair, $2N_{ch}/\langle N_{part}\rangle$ and the full width at half maximum, $FWHM$ are shown as a function of centrality for each energy for $Au+Au$ collisions. The first two quantities are shown for both the measured region $N_{ch}|_{|\eta|<5.4}$(open points) and based on extrapolations, $N_{ch}^p$ (solid circles), see text for details. Corresponding data for inelastic $\bar pp$ collisions~\protect\cite{Alner} at 200 GeV and $pp$ collisions~\protect\cite{NPB129_365} are shown as solid squares, whereas the crosses are from the present work and solid diamonds represent an overall fit $N_{ch}=-0.42+4.69s^{0.155}$ for inelastic collisions~\protect\cite{Heiselberg}. The $FWHM$ for 200 GeV $\bar pp$ and 62.4 GeV $pp$ collisions (solid squares) were obtained from Refs.~\cite{Alner} and Ref.~\cite{NPB129_365}, respectively, whereas the value (solid diamond) for 19.6 GeV represents an extrapolation based on data in the latter work. No $FWHM$ data are available for 130 GeV $pp$ collisions. Error bars and ellipses represent systematic 90\% C.L. errors.}
\label{Phmult_fig23}
\end{figure}

\begin{figure}
\centerline{\epsfig{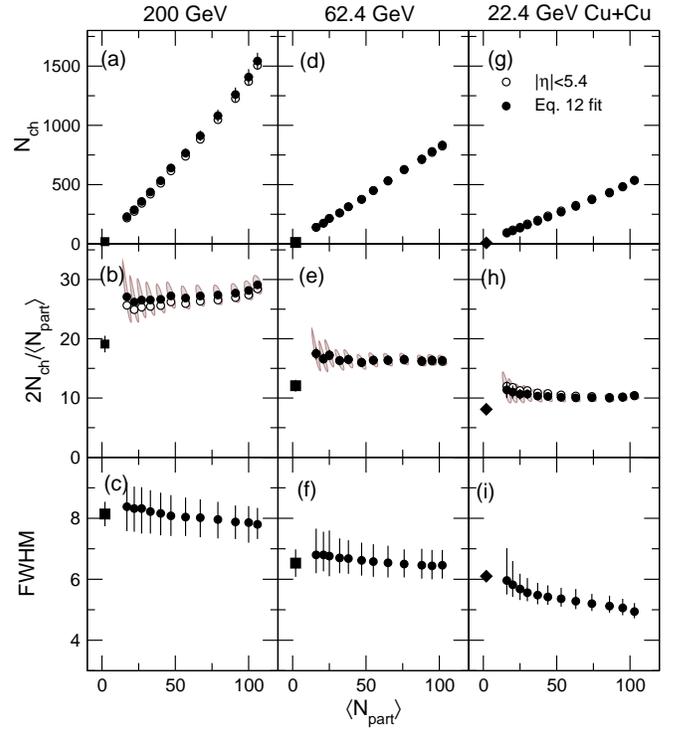}}
\caption{Same as Fig.~\ref{Phmult_fig23} but for {\it Cu+Cu} collisions. The data for 22.4 GeV $pp$ collisions (solid diamonds) were obtained as for the 19.6 GeV data shown in Fig.~\ref{Phmult_fig23}.} 
\label{Phmult_fig24}
\end{figure}

\begin{figure}
\centerline{\epsfig{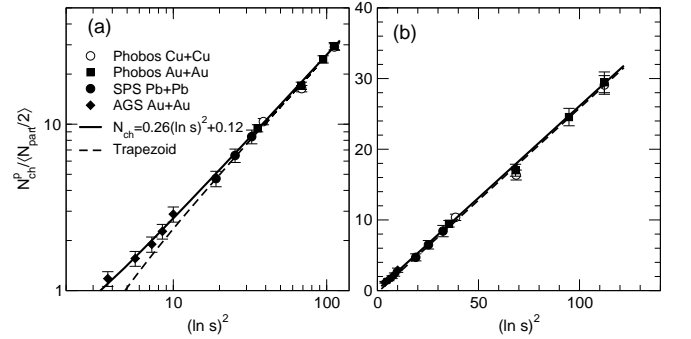}}
\caption{Doubly logarithmic (panel a) and linear (panel b) plots of $N_{ch}^p/\langle N_{part}/2 \rangle$ vs. $(\ln s)^2$. The PHOBOS results are for 0-3\% central collisions. The trend is extended to lower energies with data from the SPS~\cite{NA49} and AGS~\cite{AGS}. The dashed and solid curves were calculated using Eqs.~\ref{trap} and \ref{Nch_expression}, respectively.}
\label{Phmult_fig25}
\end{figure}

\subsubsection{Factorization}
\begin{figure}
\centerline{\epsfig{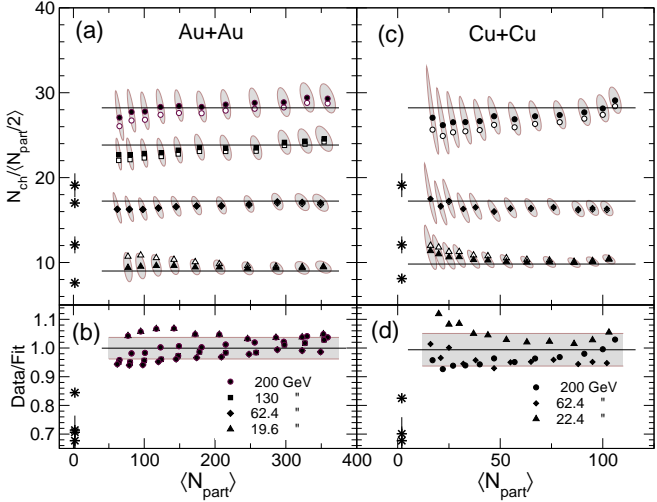}}
\caption{ Two estimates of the total charged particle multiplicity per participant pair, $2N_{ch}|_{|\eta|<5.4}/\langle N_{part}\rangle$ (open points) and $2N_{ch}^p/\langle N_{part}\rangle$ (solid points), are shown as a function of centrality for each energy for $Au+Au$ (panel a) and $Cu+Cu$ (panel c) collisions. Horizontal lines represent centrality independent levels predicted by Eq.~\ref{Nch_expression}. The lower panels show the ratio of data to the Eq.~\ref{Nch_expression} prediction. The solid line is the average of all data points and the grey-shaded area the 1$\sigma$ standard deviation. Corresponding data for $p\bar p/pp$ collisions are shown as stars, see Figs.~\ref{Phmult_fig23} and \ref{Phmult_fig24} for details.}
\label{Phmult_fig26}
\end{figure}

Similar to what was shown for the midrapidity multiplicity (Sect. IV.A.2) the total participant-scaled charged particle multiplicity, $dN_{ch}/d\eta /\langle N_{part}/2 \rangle $, appears to exhibit factorization of the centrality and energy dependence, albeit of a somewhat trivial sort since, as discussed above, no centrality dependence outside of error bars is observed. The degree to which this quantity depends only on the collision energy is illustrated in Fig.~\ref{Phmult_fig26}, where $N_{ch}^p/\langle N_{part}/2 \rangle$ (solid points with error ellipses) and $N_{ch}|_{|\eta|<5.4}/\langle N_{part}/2 \rangle$ (open circles) are compared to the predictions obtained by Eq.~\ref{Nch_expression} (solid lines). The lower panels shows the ratio between the data, $N_{ch}^p/\langle N_{part}/2 \rangle$, and the fit (Eq.~\ref{Nch_expression}). Here the solid lines are the average of all points and the grey band represents the standard deviation, which amounts to $\sigma$= 0.036 and  $\sigma$= 0.057 for $Au+Au$ and $Cu+Cu$ collisions, respectively.

\subsection{d+Au collisions}

\begin{figure}
\centerline{\epsfig{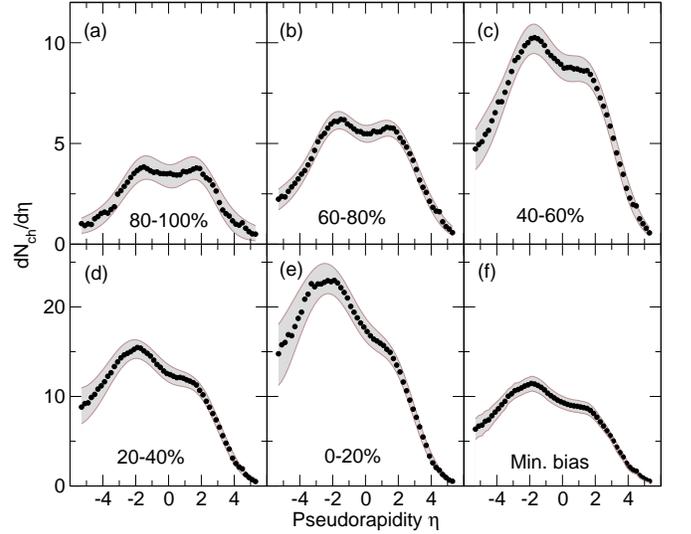}}
\caption{The charged particle multiplicity, $dN_{ch}/d\eta$, is shown for five centrality bins for $d+Au$ collisions (panels a-e). The grey-shaded area represents the 1$\sigma$ standard deviation. The minimum-bias distribution is shown in panel f. The pseudorapidity, $\eta$, is given relative to the deuteron beam, which travels in the positive $z$ direction.}
\label{Phmult_fig27}
\end{figure}

The primary reason for measuring $d+Au$ collisions in RHIC Run 3 was to obtain high $p_T$ spectra for a comparison to $Au+Au$ collisions and literature data on $pp$ collisions to determine whether the observed high $p_T$ suppression in $Au+Au$ collisions is an entrance channel effect (gluon saturation) or associated with high energy loss rates of partons traversing a color-charged medium. Concurrent with these studies, PHOBOS recorded the charged-particle multiplicity, the results of which are shown in Fig.~\ref{Phmult_fig27} for five centrality bins and a minimum-bias trigger. One observes that the $dN_{ch}/d\eta$ distribution is strongly asymmetric for the most central collisions (panel e) but approaches symmetry for the most peripheral collisions (panel a). The asymmetry is expected from momentum conservation, since a larger number of participants are associated with the incoming {\it Au} nucleus traveling in the negative $\eta$ direction, especially for central collisions, as predicted in Monte Carlo Glauber model calculations, see Table~\ref{dAu}.  The total charged-particle multiplicity, $N_{ch}|_{|\eta|<5.4}$, within the -5.4$<\eta<$5.4  acceptance region is also listed along with an estimate of the total charged particle multiplicity, $N_{ch}^{tot}$, which includes an estimate of the contribution from the unmeasured region. As detailed in Ref.~\cite{PH23} this is obtained from lower energy $p+A$ collisions by assuming extended longitudinal scaling. Also discussed in Ref.~\cite{PH23} (see Fig. 4a) is the observation that the participant-scaled total multiplicity, $N_{ch}^{tot}/\langle N_{part}/2 \rangle$, for $d+Au$ collisions is commensurate with that observed for $\overline pp$ collisions, see {\it e.g.} Fig. 13 of Ref.~\cite{Bari}. The additional ($\sim$40\%) enhancement seen for $Au+Au$ and $Cu+Cu$ systems does not appear for $d+Au$. 

\begin{table}  
\caption{Summary of the Monte Carlo Glauber model predictions of the number of participating nucleons, $\langle N_{part}^{Au}\rangle$ and $\langle N_{part}^d\rangle$ associated with the incoming $Au$ and deuteron nuclei, respectively. The total number of charged particles emitted within the PHOBOS acceptance, $N_{ch}|_{|\eta|<5.4}$ and an estimate of the total multiplicity including the unmeasured region, $N_{ch}^{tot}$ (see text for details) are also listed. Errors are 90\% C.L. systematic errors. }
\vspace{3mm}
\begin{tabular}{ccccc}
\tableline
Cent. (\%) & $\langle N_{part}^{Au}\rangle$ & $\langle N_{part}^d\rangle$ &  $N_{ch}|_{|\eta|<5.4}$ & $N_{ch}^{tot}$	\\

\hline
0-20	&	13.5$\pm$1.0	&	2.0$\pm$0.1	&	157$\pm$10	&	167$^{+14}_{-11}$\\
20-40	&	8.9$\pm$0.7	&	1.9$\pm$0.1	&	109$\pm$7	&	115$^{+10}_{-8}$ \\
40-60	&	5.4$\pm$0.6	&	1.7$\pm$0.2	&	 74$\pm$5	&	 77$^{+7}_{-5}$  \\
60-80	&	2.9$\pm$0.5	&	1.4$\pm$0.2	&	 46$\pm$3	&	 48$^{+3}_{-3}$  \\
80-100	&	1.6$\pm$0.4	&	1.1$\pm$0.2	&	 28$\pm$3	&	 29$^{+3}_{-3}$  \\
\hline
Min-Bias&	6.6$\pm$0.5	&	1.7$\pm$0.1	&	82$\pm$6	&	87$^{+7}_{-6}$\\
\tableline
\end{tabular}
\label{dAu}
\end{table}

\subsection{p+p collisions}
   
Inelastic proton-proton collisions were measured at two energies, 200 GeV and 410 GeV. The 200 GeV data were measured, in part, in order to provide a baseline against which to identify and study the special effects associated with heavy-ion collisions in observables such as charged particle multiplicities, collective flow, and high $p_T$ suppression in particle spectra. In addition, $pp$ data were collected at the highest RHIC energy, namely $\sqrt{s}$ = 410 GeV. 

Only the hit-counting method was applied in the analysis of data for pp collisions. Using HIJING Monte 
Carlo simulations, the particle yields were corrected for detector acceptance, secondary particles produced 
in the material of the detector, and particles resulting from weak decays. These corrections have similar 
dependences on $\eta$ as those found for heavy ion collisions. Pseudorapidity distributions were extracted 
in bins of the total observed charged-particle multiplicity. The distribution found for each bin was corrected 
for the effects of triggering and vertexing efficiency, which were both found to be strongly multiplicity dependent. 
The analysis compared the number of input Monte Carlo events to the number surviving all event criteria so 
separate triggering and vertex efficiencies were not extracted. Individually corrected distributions were then 
combined, weighted by the efficiency corrected number of events in each multiplicity bin, to generate the final 
average dN/d$\eta$. Systematic uncertainties, which were much larger than the statistical ones, were found by 
varying the cuts used to select events and hits, by comparing the results for positive and negative pseudorapidity, 
and by considering the differences found between the hit-counting and other techniques when applied to heavy 
ion data. Systematic uncertainties due to all sources were then added in quadrature. 

\begin{figure}
\centerline{\epsfig{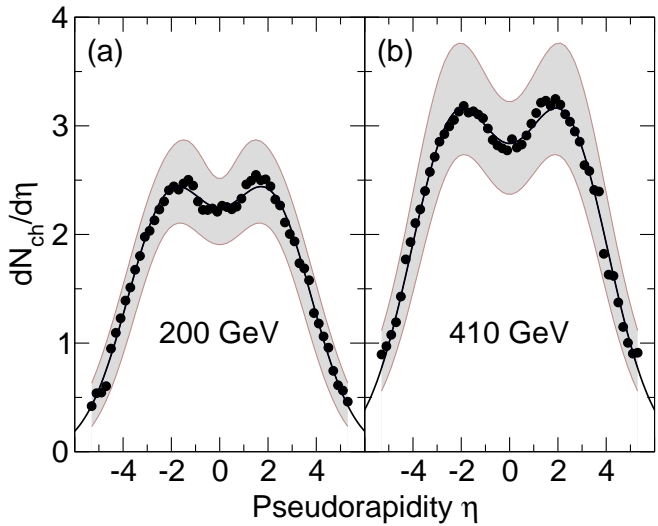}}
\caption{The charged particle multiplicity, $dN_{ch}/d\eta$, is shown for 200 GeV (panel a) and 410 GeV (panel b) for $pp$ inelastic collisions. The grey-shaded area represents the 90\% confidence limit systematic error. The solid curves are symmetric double-Gaussian fits to the data used to derive the total multiplcities listed in Table ~\ref{pp_data}. }
\label{phmult_pp}
\end{figure}

The charged-particle multiplicity distributions in pseudorapidity are presented in Fig.~\ref{phmult_pp} for 
inelastic $pp$ collisions~\cite{Sagerer_thesis}. The grey-shaded area around the data points shows the 
90\% confidence limit systematic error. The midrapidity and total charged particle multiplicity for minimum 
bias inelastic $pp$ collisions, extracted using the symmetric double-Gaussian fits to the data shown as 
solid curves in Fig.~\ref{phmult_pp}, are listed in Table~\ref{pp_data}.

\begin{table}
\caption{Table of the midrapidity charged-particle density $dN_{ch}/d\eta$, the total number of charged particles emitted within the PHOBOS acceptance, $N_{ch}|_{|\eta|<5.4}$, in minimum-bias, inelastic $p+p$ collisions.  Also listed is an estimate of the total multiplicity including the unmeasured region, $N_{ch}^{tot}$ using a three-parameter double-Gaussian fit to the data. Errors are 90\% C.L. systematic errors.}
\vspace{3mm}
\begin{tabular}{lcccc}
\tableline
$\sqrt{s}$ (GeV)& $y_{beam}$	&$\frac{dN_{ch}}{d\eta}|_{|\eta|<1}$	&	$N_{ch}|_{|\eta|<5.4}$& $N_{ch}^{tot}$	\\
\hline
200		&	5.361	& 2.25$^{+0.37}_{-0.30}$&	19.3$\pm$1.8	&	20.0$\pm$1.8	\\
410		&	6.079	& 2.87$^{+0.44}_{-0.43}$&	26.2$\pm$2.5	&	27.7$\pm$2.5	\\
\tableline
\end{tabular}
\label{pp_data}
\end{table}

\subsection{Extended longitudinal scaling}
\label{els}

\begin{figure}
\centerline{\epsfig{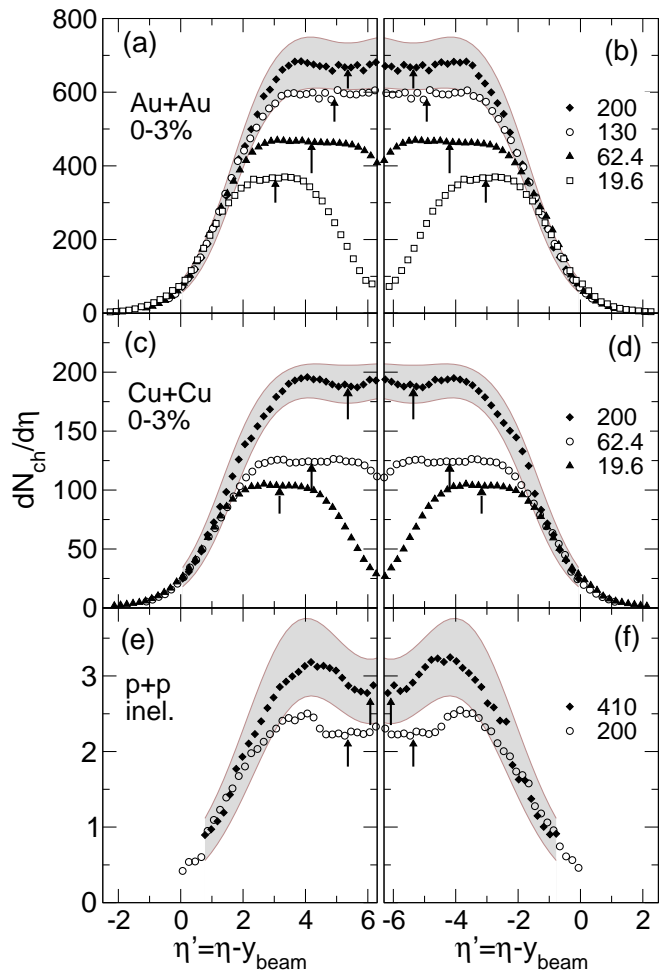}}
\caption{Illustration of the extended longitudinal scaling ({\it a.k.a.} limiting fragmentation scaling) for $Au+Au$ (panels a, b), $Cu+Cu$ (panels c, d), and $p+p$ collisions (panels e, f).  The systematic errors (90\% C.L.) are shown as shaded areas for the highest energy only for each system. The arrows indicate the location of mid-rapidity ($\eta$=0). }
\label{limfrag}
\end{figure}

It is a well-known phenomenon that, at sufficiently high energy, the particle production in the rapidity region of either collision partner becomes largely independent of the collision energy. This effect is referred to as ``limiting fragmentation scaling''\cite{Benecke} and it has been observed in $p+p, p+A$ and heavy-ion collisions~\cite{PH10}. The term ``extended longitudinal scaling'' has also been used to describe this phenomenon, since this scaling feature 
covers a more extended region of $\eta$ than expected from the hypothesis of limiting fragmentation and because it also appears to apply to other observables in heavy-ion collisions, {\it e.g.} the magnitude of elliptic flow~\cite{PH19}.  Although the original concept refers to the particle production as a function of rapidity, $y$, it can be shown that the scaling also holds for $dN_{ ch}/d\eta$. This results from the fact that for $|\eta|\gg 1$ the relation $p_T\sinh\eta = m_T\sinh y$ leads to $y \approx \ln(p_T/m_T)+\eta$, where the term $\ln(p_T/m_T)$ leads to a very small correction that is not taken into account. Similarly, the Jacobian $p_T\cosh\eta/\sqrt{m^2+p_T\cosh^2\eta}$ associated with the transformation $dN_{ch}/dy$ to $dN_{ch}/d\eta$ is close to unity in the $\eta$-region of interest and may safely be ignored. In Figure~\ref{limfrag} the extent to which the limiting fragmentation scaling is valid for $Au+Au$ (panels a,b), $Cu+Cu$ (panels c,d) and $p+p$ collisions (panels e,f) is shown in both the positive and negative $\eta$ regions. Further discussion of ``limiting fragmentation scaling'' in heavy-ion collisions can be found in Refs.~\cite{PH10,PH24}.

\section{Conclusions}
\label{Conclusions}

We have measured the multiplicity of charged particles in $Au+Au$ collisions at nucleon-nucleon center-of-mass energies of $\sqrt{s_{_{NN}}}$=19.6, 56, 62.4, 130, and 200 GeV, $Cu+Cu$ collisions at $\sqrt{s_{_{NN}}}$=22.4, 62.4, and 200 GeV, $\sqrt{s_{_{NN}}}$=200 GeV $d+Au$ collisions, and $p+p$ collisions at $\sqrt{s_{_{NN}}}$=200 and 410 GeV at the Relativistic Heavy-Ion Collider at Brookhaven National Laboratory. For the $Au+Au$ system, the complete distributions in pseudorapidity were measured at $\sqrt{s_{_{NN}}}$=19.6, 62.4, 130, and 200 GeV for 10-11 centrality bins corresponding to 45-50\% most central collisions, whereas the measurement at 56 GeV yielded only the charged particle multiplicity at midrapidity for the 6\% most central collisions. The midrapidity measurements, based on the tracklet analysis method were extended up to 70\% of the cross section in up to 15 centrality bins. For $Cu+Cu$, the pseudorapidity distributions were measured at $\sqrt{s_{_{NN}}}$=22.4, 62.4, and 200 GeV in 12 centrality bins corresponding to 55\% most central collisions. 

Measurements were also carried out for the asymmetric $d+Au$ system at $\sqrt{s_{_{NN}}}$=200 GeV, and the multiplicity distributions are reported for five centrality bins as well as minimum bias events. Also elementary $p+p$ collisions were measured at  $\sqrt{s}$=200 and 410 GeV and the results given in the present work.

The multiplicity results were in all cases derived from three different methods of analysis, which all agree within systematic errors. The multiplicities were all corrected for missing low $p_T$ particles and the contribution from weak decays have been subtracted.
At midrapidity, we observe a smooth decrease of charged particle production per participant nucleon pair when going to more peripheral collisions trending towards the results from $pp$ and $\overline pp$ collisions for the most peripheral collisions at the same energies. Surprisingly, the midrapidity production can be factorized into separate energy and centrality dependences for both the $Au+Au$ and $Cu+Cu$ data. Combined with lower energy data from the SPS (CERN) and AGS (Brookhaven) we find an approximate logarithmic energy dependence of the midrapidity charged particle production.  

A unique feature of the PHOBOS experiment is that the pseudorapidity distributions were measured over a very wide range of $-5.4<\eta<5.4$. Three estimates of the total charged particle multiplicity in $Au+Au$ and $Cu+Cu$ collisions were performed. One method attempts to exclude the contribution from excited spectators, which appear to contribute substantially at the highest pseudorapidities for the lowest collision energies, by using an analytical function fit to the data within the pseudorapidity range $-y_{beam} < \eta < y_{beam}$. By extrapolation into the unmeasured $\eta$-region, of the order 1-5\% in total multiplicity, it is thus possible to estimate the total charged particle multiplicity generated in participant collisions. When scaled to the number of participant pairs, this multiplicity is constant as a function of centrality and about 40\% larger than those observed for $pp$ and $\overline pp$ collisions. Approximating the central pseudorapidity distributions by a trapezoidal shape, the height of which is given by the logarithmic midrapidity dependence, and the width is derived from the fact that the wings of the distribution follow the extended longitudinal scaling, we have found that the total charged particle production follows a simple $(\ln s)^2$ scaling, which applies over the full energy range from the lowest AGS energies $\sqrt{s}$=2.4 GeV to the most energetic RHIC collisions $\sqrt{s}$=200 GeV. However, an interesting and statistically significant departure from the otherwise smooth energy dependence of the total multiplicity and midrapidity density for all centralities is seen for $\sqrt{s_{_{NN}}} \approx$60 GeV, which may be an indication of curvature.

The widths of the pseudorapidity distributions are found to decrease with centrality at a rate that is approximately compensated for by the increase in the midrapidity plateau with centrality such that the total multiplicity is found to be almost constant when normalized to the number of participant pairs $\langle N_{\it part} \rangle/2$.

Finally, it should be noted that most of the prominent trends seen in the PHOBOS multiplicity data are not specific to $A+A$ collisions at RHIC energies. For example, $N_{part}$ scaling is seen in all $hadron + A$ collisions \cite{Busza,refA}, extended longitudinal scaling is seen in $p+p$ and $e^+e^-$ collisions \cite{PH24}, and also in $p+A$ collisions \cite{refA,refE}, and the logarithmic rise of the midrapidity particle density is seen in $A+A$ collisions at lower energies (see Fig.~\ref{Phmult_fig12}).

A complete tabulation of the pseudorapidity distribution data shown in Figs.~\ref{Phmult_fig16}-\ref{Phmult_fig22},~\ref{Phmult_fig27}, and \ref{phmult_pp} as well as various derived quantities not listed in this article, may be obtained from the Electronic Physics Auxiliary Publication Service of the American Institute of Physics. See EPAPS Document No. [number will be inserted by publisher ]. A direct link to this document may be found in the online article's HTML reference section. The document may also be reached via the EPAPS homepage (http://www.aip.org/pubservs/epaps.html) or from ftp.aip.org in the directory /epaps/. See the EPAPS homepage for more information.

\section*{Acknowledgments}
%
%
%
%
This work was partially supported by U.S. DOE grants 
DE-AC02-98CH10886,
DE-FG02-93ER40802, 
DE-FG02-94ER40818,  
DE-FG02-94ER40865, 
DE-FG02-99ER41099, and
DE-AC02-06CH11357, by U.S. 
NSF grants 9603486, 
0072204,            
and 0245011,        
by Polish MNiSW grant N N202 282234 (2008-2010),
by NSC of Taiwan Contract NSC 89-2112-M-008-024, and
by Hungarian OTKA grant (F 049823).

\section*{Appendix: Au+Au and Cu+Cu multiplicity tables}
  
In this appendix we give the tables of relevant parameters for the PHOBOS charged particle multiplicity data for $Au+Au$ and $Cu+Cu$ collisions. The corresponding parameters for $d+Au$ and $p+p$ collisions are given in tables~\ref{dAu} and \ref{pp_data} in the main text.

\begin{table}[h] 
\caption{Summary of the midrapidity $\frac{dN_{\it ch}}{d\eta}|_{|\eta|<1}$ charged particle multiplicity for Au+Au collisions obtained from the tracklet analysis. The data are listed as a function of centrality expressed in percentage of the total reaction cross section for all four energies. Columns 2 and 4 list derived quantities, namely number of participants as well as the midrapidity density normalized to the number of participant pairs $N_{\it part}/2$. The errors are systematic errors at 90\% C. L.; statistical errors are negligible. Note, the table continues overleaf.}
\vspace{3mm}
\begin{tabular}{cccc}
\multicolumn{4}{c}{Au+Au $\sqrt{s_{_{NN}}}$=19.6 GeV} $y_{\it beam}$=3.036\\
\hline
  Bin & $N_{\it part}$ & $\frac{dN}{d\eta}|_{|\eta|<1}$  &  $\frac{dN_{\it ch}/d\eta|_{|\eta<1|}}{N_{\it part}/2}$ \\
\hline
0-3\%    & 351$\pm$11  & 331$\pm$24   & 1.89$\pm$0.15 \\
3-6\%    & 322$\pm$10  & 297$\pm$22   & 1.84$\pm$0.15 \\
6-10\%   & 286$\pm$9   & 260$\pm$20   & 1.82$\pm$0.15 \\
10-15\%  & 247$\pm$8   & 216$\pm$16   & 1.76$\pm$0.14 \\
15-20\%  & 206$\pm$8   & 181$\pm$14   & 1.75$\pm$0.15 \\
20-25\%  & 171$\pm$7   & 148$\pm$11   & 1.73$\pm$0.15 \\
25-30\%  & 142$\pm$7   & 121$\pm$9    & 1.70$\pm$0.15 \\
30-35\%  & 117$\pm$7   &  97$\pm$7    & 1.65$\pm$0.16 \\
35-40\%  &  95$\pm$7   &  78$\pm$6    & 1.64$\pm$0.17 \\
40-45\%  &  74$\pm$6   &  59$\pm$4    & 1.61$\pm$0.18 \\
45-50\%  &  58$\pm$3   &  45$\pm$3    & 1.56$\pm$0.19 \\
50-55\%  &  45$\pm$3   &  35$\pm$3    & 1.55$\pm$0.20 \\
55-60\%  &  34$\pm$3   &  26$\pm$2    & 1.55$\pm$0.22 \\
\hline
\multicolumn{4}{c}{Au+Au $\sqrt{s_{_{NN}}}$=62.4 GeV} $y_{\it beam}$=4.196\\
\hline
  Bin &  $N_{\it part}$ & $\frac{dN}{d\eta}|_{|\eta|<1}$  & $\frac{dN_{\it ch}/d\eta|_{|\eta<1|}}{N_{\it part}/2}$ \\
\hline
0-3\%   & 356$\pm$11 & 492$\pm$36   & 2.76$\pm$0.23 \\
3-6\%   & 325$\pm$10 & 433$\pm$32   & 2.67$\pm$0.22 \\
6-10\%  & 288$\pm$9  & 377$\pm$28   & 2.62$\pm$0.21 \\
10-15\% & 248$\pm$8  & 316$\pm$23   & 2.55$\pm$0.21 \\
15-20\% & 209$\pm$7  & 260$\pm$19   & 2.50$\pm$0.20 \\
20-25\% & 174$\pm$7  & 212$\pm$15   & 2.44$\pm$0.21 \\
25-30\% & 145$\pm$7  & 174$\pm$13   & 2.41$\pm$0.21 \\
30-35\% & 119$\pm$7  & 140$\pm$10   & 2.35$\pm$0.22 \\
35-40\% &  98$\pm$7  & 111$\pm$8    & 2.28$\pm$0.23 \\
40-45\% &  78$\pm$6  &  87$\pm$6    & 2.24$\pm$0.25 \\
45-50\% &  62$\pm$6  &  67$\pm$5    & 2.16$\pm$0.26 \\
50-55\% &  48$\pm$5  &  50$\pm$4    & 2.10$\pm$0.27 \\
50-60\% &  36$\pm$4  &  36$\pm$3    & 2.01$\pm$0.28 \\
60-65\% &  27$\pm$3  &  25$\pm$2    & 1.91$\pm$0.28 \\
65-70\% &  19$\pm$3  &  17$\pm$1    & 1.77$\pm$0.27 \\
\hline
\end{tabular}
\label{table1}
\end{table}

\clearpage

\setcounter{table}{4}
\begin{table}  
\caption{Continued.}

\vspace{3mm}
\begin{tabular}{cccc}
\multicolumn{4}{c}{Au+Au $\sqrt{s_{_{NN}}}$=130 GeV} $y_{\it beam}$=4.930\\
\hline
  Bin & $N_{\it part}$ & $\frac{dN}{d\eta}|_{|\eta|<1}$  & $\frac{dN_{\it ch}/d\eta|_{|\eta<1|}}{N_{\it part}/2}$ \\
\hline
0-3\%   & 355$\pm$12 & 613$\pm$24   & 3.45$\pm$0.17 \\
3-6\%   & 330$\pm$10 & 545$\pm$21   & 3.31$\pm$0.16 \\
6-10\%  & 295$\pm$9  & 472$\pm$18   & 3.20$\pm$0.16 \\
10-15\% & 254$\pm$8  & 393$\pm$15   & 3.09$\pm$0.16 \\
15-20\% & 214$\pm$8  & 327$\pm$13   & 3.06$\pm$0.16 \\
20-25\% & 179$\pm$7  & 274$\pm$11   & 3.06$\pm$0.17 \\
25-30\% & 148$\pm$6  & 220$\pm$8    & 2.96$\pm$0.17 \\
30-35\% & 122$\pm$6  & 180$\pm$7    & 2.94$\pm$0.18 \\
35-40\% & 100$\pm$5  & 140$\pm$5    & 2.80$\pm$0.18 \\
40-45\% &  80$\pm$5  & 110$\pm$4    & 2.75$\pm$0.20 \\
45-50\% &  63$\pm$4  &  83$\pm$3    & 2.64$\pm$0.21 \\
\hline
\hline
\multicolumn{4}{c}{Au+Au $\sqrt{s_{_{NN}}}$=200 GeV} $y_{\it beam}$=5.361\\
\hline
  Bin & $N_{\it part}$& $\frac{dN}{d\eta}|_{|\eta|<1}$  & $\frac{dN_{\it ch}/d\eta|_{|\eta<1|}}{N_{\it part}/2}$ \\
\hline
0-3\%   & 361$\pm$11 & 691$\pm$52   & 3.82$\pm$0.31 \\
3-6\%   & 331$\pm$10 & 619$\pm$46   & 3.74$\pm$0.30 \\
6-10\%  & 297$\pm$9  & 540$\pm$41   & 3.64$\pm$0.30 \\
10-15\% & 255$\pm$8  & 465$\pm$35   & 3.65$\pm$0.30 \\
15-20\% & 215$\pm$7  & 384$\pm$29   & 3.57$\pm$0.29 \\
20-25\% & 180$\pm$7  & 313$\pm$24   & 3.47$\pm$0.30 \\
25-30\% & 150$\pm$6  & 257$\pm$19   & 3.42$\pm$0.29 \\
30-35\% & 124$\pm$6  & 208$\pm$16   & 3.37$\pm$0.30 \\
35-40\% & 101$\pm$6  & 165$\pm$12   & 3.25$\pm$0.31 \\
40-45\% &  82$\pm$6  & 133$\pm$10   & 3.25$\pm$0.34 \\
45-50\% &  65$\pm$6  & 100$\pm$8    & 3.10$\pm$0.38 \\
50-55\% &  49$\pm$5  &  73$\pm$5    & 2.98$\pm$0.37 \\
55-60\% &  37$\pm$4  &  54$\pm$4    & 2.88$\pm$0.39 \\
60-65\% &  28$\pm$3  &  38$\pm$3    & 2.78$\pm$0.40 \\
65-70\% &  20$\pm$3  &  27$\pm$2    & 2.68$\pm$0.41 \\
\hline
\end{tabular}
\label{table1a}
\end{table}

\begin{table}  
\caption{Same as Table~\ref{table1}, but for Cu+Cu collisions.}
\vspace{3mm}
\begin{tabular}{cccc}
\multicolumn{4}{c}{Cu+Cu $\sqrt{s_{_{NN}}}$=22.4 GeV} $y_{\it beam}$=3.170\\
\hline
  Bin & $N_{\it part}$& $\frac{dN}{d\eta}|_{|\eta|<1}$  & $\frac{dN_{\it ch}/d\eta|_{|\eta<1|}}{N_{\it part}/2}$\\
\hline
0-3\%   & 103$\pm$ 3 & 101$\pm$8 & 1.96$\pm$0.16 \\
3-6\%   &  95$\pm$ 3 &  90$\pm$7 & 1.90$\pm$0.15 \\
6-10\%  &  86$\pm$ 3 &  80$\pm$6 & 1.86$\pm$0.15 \\
10-15\% &  74$\pm$ 3 &  68$\pm$5 & 1.83$\pm$0.15 \\
15-20\% &  63$\pm$ 3 &  57$\pm$4 & 1.82$\pm$0.16 \\
20-25\% &  53$\pm$ 3 &  48$\pm$4 & 1.81$\pm$0.17 \\
25-30\% &  44$\pm$ 3 &  40$\pm$3 & 1.80$\pm$0.19 \\
30-35\% &  37$\pm$ 3 &  33$\pm$2 & 1.78$\pm$0.21 \\
35-40\% &  30$\pm$ 3 &  27$\pm$2 & 1.79$\pm$0.24 \\
40-45\% &  24$\pm$ 3 &  22$\pm$2 & 1.76$\pm$0.26 \\
45-50\% &  20$\pm$ 3 &  18$\pm$1 & 1.77$\pm$0.29 \\
\hline
\multicolumn{4}{c}{Cu+Cu $\sqrt{s_{_{NN}}}$=62.4 GeV} $y_{\it beam}$=4.196\\
\hline
  Bin & $N_{\it part}$& $\frac{dN}{d\eta}|_{|\eta|<1}$  & $\frac{dN_{\it ch}/d\eta|_{|\eta<1|}}{N_{\it part}/2}$\\
\hline
0-3\%   &106$\pm$3 & 138$\pm$10 & 2.64$\pm$0.21 \\
3-6\%   & 97$\pm$3 & 123$\pm$ 9 & 2.55$\pm$0.21 \\
6-10\%  & 88$\pm$3 & 108$\pm$ 8 & 2.46$\pm$0.20 \\
10-15\% & 76$\pm$3 &  92$\pm$ 7 & 2.40$\pm$0.20 \\
15-20\% & 65$\pm$3 &  77$\pm$ 6 & 2.37$\pm$0.20 \\
20-25\% & 55$\pm$3 &  64$\pm$ 5 & 2.33$\pm$0.22 \\
25-30\% & 47$\pm$3 &  52$\pm$ 4 & 2.25$\pm$0.23 \\
30-35\% & 38$\pm$3 &  43$\pm$ 3 & 2.26$\pm$0.26 \\
35-40\% & 32$\pm$3 &  35$\pm$ 3 & 2.22$\pm$0.28 \\
40-45\% & 26$\pm$3 &  28$\pm$ 2 & 2.21$\pm$0.32 \\
45-50\% & 21$\pm$3 &  23$\pm$ 2 & 2.21$\pm$0.35 \\
\hline
\multicolumn{4}{c}{Cu+Cu $\sqrt{s_{_{NN}}}$=200 GeV} $y_{\it beam}$=5.361\\
\hline
  Bin & $N_{\it part}$& $\frac{dN}{d\eta}|_{|\eta|<1}$  & $\frac{dN_{\it ch}/d\eta|_{|\eta<1|}}{N_{\it part}/2}$\\
\hline
0-3\%   & 108$\pm$4 & 198$\pm$15  & 3.66$\pm$0.29  \\
3-6\%   & 101$\pm$3 & 175$\pm$13  & 3.48$\pm$0.28  \\
6-10\%  &  91$\pm$3& 155$\pm$12   & 3.42$\pm$0.28  \\
10-15\% &  79$\pm$3 & 132$\pm$10  & 3.33$\pm$0.27  \\
15-20\% &  67$\pm$3 & 109$\pm$ 8  & 3.26$\pm$0.28  \\
20-25\% &  57$\pm$3 &  91$\pm$ 7  & 3.21$\pm$0.29  \\
25-30\% &  48$\pm$3 &  75$\pm$ 6  & 3.17$\pm$0.32  \\
30-35\% &  40$\pm$3 &  62$\pm$ 5  & 3.15$\pm$0.35  \\
35-40\% &  33$\pm$3 &  50$\pm$ 4  & 3.07$\pm$0.38  \\
40-45\% &  27$\pm$3 &  40$\pm$ 3  & 3.04$\pm$0.43  \\
45-50\% &  22$\pm$3 &  32$\pm$ 2  & 2.97$\pm$0.46  \\
\hline
\end{tabular}
\label{table2}
\end{table}

\begin{table}  
\caption{Summary of the total charged particle multiplicity estimates for Au+Au collisions obtained in three different ways. $N_{\it ch}|_{|\eta|<5.4}$ denotes the multiplicity observed within the acceptance region $|\eta|<5.4$ obtained from the single layer analysis, whereas $N_{\it ch}^{tot}$ and  $N_{\it ch}^p$ represent extrapolations into the unmeasured $\eta$ region, either including or excluding contributions from spectator emission. See text for details. The data are listed as a function of centrality expressed in percentage of the total reaction cross section for all four energies. Also listed are the full width at half maximum, FWHM, of the $dN/d\eta$ distributions and the derived quantity $N_{\it part}$. Only systematic errors at 90\% C. L. are given since statistical errors are negligible.}
\vspace{3mm}
\begin{tabular}{cccccc}
\multicolumn{6}{c}{Au+Au $\sqrt{s_{_{NN}}}$=19.6 GeV} $y_{\it beam}$=3.036\\
\hline
  Bin & $N_{\it part}$& FWHM & $N_{\it ch}|_{|\eta|<5.4}$ & $N_{\it ch}^{tot}$ & $N_{\it ch}^p$ \\
\hline
0-3\%   & 353$\pm$11  & 4.80$^{+0.2}_{-0.2}$ &  1682$\pm$115 & 1676$\pm$115 & 1669$\pm$115\\
3-6\%   & 323$\pm$10  & 4.86$^{+0.2}_{-0.3}$ &  1531$\pm$111 & 1522$\pm$111 & 1512$\pm$111\\
6-10\%  & 286$\pm$9   & 4.94$^{+0.3}_{-0.3}$ &  1367$\pm$97  & 1352$\pm$97  & 1337$\pm$97\\
10-15\% & 246$\pm$8   & 5.03$^{+0.3}_{-0.3}$ &  1182$\pm$86 & 1162$\pm$86  & 1145$\pm$86\\
15-20\% & 206$\pm$8   & 5.10$^{+0.3}_{-0.3}$ &  1014$\pm$78 &  989$\pm$78  & 973$\pm$78\\
20-25\% & 172$\pm$7   & 5.19$^{+0.3}_{-0.3}$ &   866$\pm$71 &  836$\pm$71  & 813$\pm$71\\
25-30\% & 142$\pm$7   & 5.28$^{+0.4}_{-0.3}$ &   735$\pm$68 &  707$\pm$68  & 684$\pm$68\\
30-35\% & 117$\pm$7   & 5.37$^{+0.4}_{-0.3}$ &   617$\pm$66 &  586$\pm$66  & 563$\pm$66\\
35-40\% &  95$\pm$7   & 5.41$^{+0.5}_{-0.2}$ &   516$\pm$63 &  482$\pm$63  & 453$\pm$63\\
40-45\% &  -          & -                    &   418$\pm$60 &  387$\pm$60    & 362$\pm$60\\

\hline
\multicolumn{6}{c}{Au+Au $\sqrt{s_{_{NN}}}$=62.4 GeV} $y_{\it beam}$=4.196\\
\hline
  Bin & $N_{\it part}$& FWHM & $N_{\it ch}|_{|\eta|<5.4}$ & $N_{\it ch}^{tot}$ & $N_{\it ch}^p$ \\
\hline
0-3\%   & 349$\pm$11  &  6.47$^{+0.3}_{-0.4}$ & 2935$\pm$147  & 2988$\pm$149 & 2971$\pm$149 \\
3-6\%   & 323$\pm$10  &  6.49$^{+0.4}_{-0.4}$ & 2733$\pm$137  & 2775$\pm$138 & 2762$\pm$138\\
6-10\%  & 288$\pm$8   &  6.54$^{+0.3}_{-0.4}$ & 2448$\pm$122  & 2489$\pm$124 & 2471$\pm$124\\
10-15\% & 248$\pm$7   &  6.62$^{+0.4}_{-0.4}$ & 2077$\pm$103  & 2120$\pm$106 & 2094$\pm$106\\
15-20\% & 209$\pm$7   &  6.68$^{+0.4}_{-0.4}$ & 1739$\pm$87   & 1777$\pm$88  & 1747$\pm$88\\
20-25\% & 174$\pm$7   &  6.74$^{+0.4}_{-0.4}$ & 1448$\pm$72   & 1485$\pm$74  & 1451$\pm$74\\
25-30\% & 145$\pm$7   &  6.83$^{+0.4}_{-0.4}$ & 1200$\pm$60   & 1236$\pm$61  & 1203$\pm$61\\
30-35\% & 120$\pm$7   &  6.91$^{+0.4}_{-0.7}$ &  984$\pm$49   & 1027$\pm$51  & 986$\pm$51\\
35-40\% &  98$\pm$6   &  6.96$^{+0.4}_{-0.5}$ &  797$\pm$40   &  840$\pm$42  & 796$\pm$42\\
40-45\% &  79$\pm$6   &  7.05$^{+0.5}_{-0.5}$ &  644$\pm$32   &  679$\pm$33  & 641$\pm$33\\
45-50\% &  62$\pm$6   &  7.08$^{+0.5}_{-0.5}$ &  504$\pm$25   &  532$\pm$26  & 505$\pm$26\\
\hline
\multicolumn{6}{c}{Au+Au $\sqrt{s_{_{NN}}}$=130 GeV} $y_{\it beam}$=4.930\\
\hline
  Bin & $N_{\it part}$& FWHM & $N_{\it ch}|_{|\eta|<5.4}$ & $N_{\it ch}^{tot}$ & $N_{\it ch}^p$ \\
\hline

0-3\%   & 354$\pm$12  &  7.15$^{+0.3}_{-0.3}$ & 4286$\pm$214  & 4376$\pm$219 & 4346$\pm$219\\
3-6\%   & 327$\pm$10  &  7.21$^{+0.3}_{-0.3}$ & 3915$\pm$196  & 4015$\pm$201 & 3971$\pm$201\\
6-10\%  & 298$\pm$9   &  7.29$^{+0.3}_{-0.3}$ & 3546$\pm$182  & 3649$\pm$182 & 3598$\pm$182\\
10-15\% & 258$\pm$8   &  7.33$^{+0.3}_{-0.3}$ & 2982$\pm$149  & 3090$\pm$155 & 3032$\pm$155\\
15-20\% & 215$\pm$8   &  7.39$^{+0.3}_{-0.3}$ & 2482$\pm$124  & 2586$\pm$129 & 2525$\pm$129\\
20-25\% & 178$\pm$7   &  7.46$^{+0.4}_{-0.4}$ & 2056$\pm$103  & 2164$\pm$108 & 2096$\pm$108\\
25-30\% & 148$\pm$6   &  7.52$^{+0.4}_{-0.4}$ & 1686$\pm$84   & 1793$\pm$90  & 1719$\pm$90\\
30-35\% & 124$\pm$6   &  7.60$^{+0.4}_{-0.4}$ & 1395$\pm$70   & 1502$\pm$75  & 1422$\pm$75\\
35-40\% & 100$\pm$5   &  7.65$^{+0.4}_{-0.4}$ & 1116$\pm$56   & 1222$\pm$61  & 1142$\pm$61\\
40-45\% &  80$\pm$5   &  7.74$^{+0.5}_{-0.5}$ &  885$\pm$44   &  975$\pm$49  &  908$\pm$49\\
45-50\% &  64$\pm$4   &  7.81$^{+0.4}_{-0.4}$ &  705$\pm$35   &  782$\pm$39  &  726$\pm$39\\
\hline
\multicolumn{6}{c}{Au+Au $\sqrt{s_{_{NN}}}$=200 GeV} $y_{\it beam}$=5.361\\
\hline
  Bin & $N_{\it part}$& FWHM & $N_{\it ch}|_{|\eta|<5.4}$ & $N_{\it ch}^{tot}$ & $N_{\it ch}^p$ \\
\hline

0-3\%   & 359$\pm$11  &   7.63$^{+0.3}_{-0.4}$ & 5159$\pm$258 & 5290$\pm$264& 5261$\pm$264 \\
3-6\%   & 330$\pm$10  &   7.64$^{+0.3}_{-0.4}$ & 4753$\pm$238 & 4895$\pm$245& 4854$\pm$245 \\
6-10\%  & 297$\pm$9   &   7.71$^{+0.4}_{-0.4}$ & 4198$\pm$210 & 4341$\pm$217& 4288$\pm$217 \\
10-15\% & 256$\pm$8   &   7.77$^{+0.4}_{-0.4}$ & 3598$\pm$180 & 3763$\pm$188& 3687$\pm$188 \\
15-20\% & 215$\pm$7   &   7.83$^{+0.4}_{-0.4}$ & 2999$\pm$150 & 3153$\pm$158& 3075$\pm$158 \\
20-25\% & 181$\pm$7   &   7.90$^{+0.4}_{-0.4}$ & 2496$\pm$125 & 2645$\pm$132& 2564$\pm$132 \\
25-30\% & 149$\pm$6   &   7.99$^{+0.4}_{-0.4}$ & 2057$\pm$103 & 2184$\pm$109& 2119$\pm$109 \\
30-35\% & 123$\pm$6   &   7.96$^{+0.5}_{-0.4}$ & 1685$\pm$84  & 1819$\pm$91 & 1742$\pm$91 \\
35-40\% & 101$\pm$6   &   8.04$^{+0.5}_{-0.5}$ & 1354$\pm$68 & 1486$\pm$74 & 1403$\pm$74 \\
40-45\% &  82$\pm$6   &   8.11$^{+0.4}_{-0.5}$ & 1096$\pm$55 & 1204$\pm$60 & 1137$\pm$60 \\
45-50\% &  65$\pm$6   &   8.23$^{+0.5}_{-0.5}$ &  847$\pm$42 &  951$\pm$48 &  880$\pm$48 \\
\hline
\end{tabular}
\label{table3}
\end{table}

\begin{table}  
\caption{Same as Table~\ref{table3} but for Cu+Cu collisions.}
\vspace{3mm}
\begin{tabular}{cccccc}
\multicolumn{6}{c}{Cu+Cu $\sqrt{s_{_{NN}}}$=22.4 GeV} $y_{\it beam}$=3.170\\
\hline
  Bin & $N_{\it part}$& FWHM & $N_{\it ch}|_{|\eta|<5.4}$ & $N_{\it ch}^{tot}$ & $N_{\it ch}^p$ \\
\hline
0-3\%   & 103$\pm$3  & 5.96$^{+1.06}_{-0.46}$ &  538$\pm$22 & 535$\pm$23 & 534$\pm$23\\
3-6\%   &  95$\pm$3  & 5.82$^{+0.78}_{-0.40}$ &  485$\pm$20 & 482$\pm$21 & 480$\pm$21\\
6-10\%  &  86$\pm$3  & 5.68$^{+0.50}_{-0.36}$ &  435$\pm$18 & 431$\pm$19 & 429$\pm$19\\
10-15\% &  74$\pm$3  & 5.56$^{+0.48}_{-0.28}$ &  380$\pm$16 & 375$\pm$18 & 372$\pm$18\\
15-20\% &  63$\pm$3  & 5.48$^{+0.40}_{-0.30}$ &  326$\pm$14 & 320$\pm$15 & 316$\pm$15\\
20-25\% &  53$\pm$3  & 5.42$^{+0.38}_{-0.26}$ &  279$\pm$12 & 273$\pm$14 & 268$\pm$14\\
25-30\% &  44$\pm$3  & 5.36$^{+0.36}_{-0.26}$ &  237$\pm$10 & 230$\pm$13 & 226$\pm$13\\
30-35\% &  37$\pm$3  & 5.28$^{+0.40}_{-0.26}$ &  201$\pm$ 9 & 194$\pm$12 & 191$\pm$12\\
35-40\% &  30$\pm$3  & 5.20$^{+0.32}_{-0.24}$ &  169$\pm$ 8 & 162$\pm$12 & 160$\pm$12\\
40-45\% &  25$\pm$3  & 5.12$^{+0.34}_{-0.22}$ &  141$\pm$ 7 & 135$\pm$11 & 133$\pm$11\\
45-50\% &  20$\pm$3  & 5.06$^{+0.30}_{-0.24}$ &  118$\pm$ 6 & 112$\pm$11 & 110$\pm$11\\
50-55\% &  16$\pm$3  & 4.94$^{+0.28}_{-0.22}$ &   96$\pm$ 6 &  92$\pm$11 &  91$\pm$11\\

\hline
\multicolumn{6}{c}{Cu+Cu $\sqrt{s_{_{NN}}}$=62.4 GeV} $y_{\it beam}$=4.196\\
\hline
  Bin & $N_{\it part}$& FWHM & $N_{\it ch}|_{|\eta|<5.4}$ & $N_{\it ch}^{tot}$ & $N_{\it ch}^p$ \\
\hline
0-3\%   & 102$\pm$3  &  6.46$^{+0.50}_{-0.55}$ & 824$\pm$36 & 833$\pm$36 & 834$\pm$36\\
3-6\%   &  95$\pm$3  &  6.44$^{+0.52}_{-0.44}$ & 771$\pm$34 & 781$\pm$34 & 780$\pm$34\\
6-10\%  &  88$\pm$3  &  6.46$^{+0.54}_{-0.44}$ & 710$\pm$31 & 721$\pm$32 & 717$\pm$32\\
10-15\% &  76$\pm$3  &  6.50$^{+0.48}_{-0.44}$ & 624$\pm$27 & 635$\pm$27 & 629$\pm$27\\
15-20\% &  65$\pm$3  &  6.54$^{+0.56}_{-0.48}$ & 530$\pm$24 & 541$\pm$24 & 534$\pm$24\\
20-25\% &  55$\pm$3  &  6.58$^{+0.57}_{-0.50}$ & 449$\pm$20 & 460$\pm$21 & 451$\pm$21\\
25-30\% &  47$\pm$3  &  6.62$^{+0.58}_{-0.50}$ & 375$\pm$17 & 386$\pm$17 & 377$\pm$17\\
30-35\% &  38$\pm$3  &  6.68$^{+0.60}_{-0.50}$ & 313$\pm$14 & 323$\pm$15 & 314$\pm$15\\
35-40\% &  32$\pm$3  &  6.70$^{+0.64}_{-0.50}$ & 261$\pm$12 & 270$\pm$13 & 261$\pm$13\\
40-45\% &  25$\pm$3  &  6.76$^{+0.84}_{-0.66}$ & 214$\pm$11 & 223$\pm$11 & 216$\pm$11\\
45-50\% &  21$\pm$3  &  6.80$^{+0.76}_{-0.54}$ & 174$\pm$ 9 & 183$\pm$ 9 & 175$\pm$ 9\\
50-55\% &  16$\pm$3  &  6.80$^{+0.86}_{-0.60}$ & 140$\pm$ 7 & 147$\pm$ 8 & 140$\pm$ 8\\    
\hline
\multicolumn{6}{c}{Cu+Cu $\sqrt{s_{_{NN}}}$=200 GeV} $y_{\it beam}$=5.361\\
\hline
  Bin & $N_{\it part}$& FWHM & $N_{\it ch}|_{|\eta|<5.4}$ & $N_{\it ch}^{tot}$ & $N_{\it ch}^p$ \\
\hline
0-3\%   & 106$\pm$3  &   7.80$^{+0.54}_{-0.48}$ & 1506$\pm$67 & 1541$\pm$70 & 1542$\pm$70\\
3-6\%   & 100$\pm$3  &   7.86$^{+0.54}_{-0.66}$ & 1370$\pm$66 & 1407$\pm$68 & 1407$\pm$68\\
6-10\%  &  91$\pm$3  &   7.88$^{+0.54}_{-0.56}$ & 1226$\pm$57 & 1262$\pm$59 & 1260$\pm$59\\
10-15\% &  79$\pm$3  &   7.96$^{+0.58}_{-0.62}$ & 1048$\pm$49 & 1084$\pm$51 & 1081$\pm$51\\
15-20\% &  67$\pm$3  &   8.02$^{+0.60}_{-0.64}$ &  882$\pm$42 &  917$\pm$43 &  912$\pm$43\\
20-25\% &  57$\pm$3  &   8.04$^{+0.64}_{-0.62}$ &  739$\pm$35 &  771$\pm$38 &  766$\pm$38\\
25-30\% &  47$\pm$3  &   8.08$^{+0.68}_{-0.66}$ &  616$\pm$30 &  645$\pm$32 &  640$\pm$32\\
30-35\% &  40$\pm$3  &   8.16$^{+0.68}_{-0.72}$ &  512$\pm$25 &  538$\pm$27 &  533$\pm$27\\
35-40\% &  33$\pm$3  &   8.22$^{+0.70}_{-0.72}$ &  420$\pm$21 &  444$\pm$23 &  438$\pm$23\\
40-45\% &  27$\pm$3  &   8.32$^{+0.70}_{-0.82}$ &  342$\pm$17 &  364$\pm$19 &  358$\pm$19\\
45-50\% &  22$\pm$3  &   8.32$^{+0.72}_{-0.76}$ &  274$\pm$14 &  293$\pm$15 &  288$\pm$15\\
50-55\% &  17$\pm$3  &   8.38$^{+0.80}_{-0.80}$ &  218$\pm$12 &  234$\pm$13 &  230$\pm$13\\
\hline
\end{tabular}
\label{table4}
\end{table}



\clearpage
\begin{references}

\bibitem{PhobosNIM} B. B. Back {\it et al.} (PHOBOS), Nucl. Instr. Meth. {\bf A499}, 603 (2003)  

\bibitem{PH1} B. B. Back {\it et al.} (PHOBOS), Phys. Rev. Lett. {\bf 85}, 3100 (2000)

\bibitem{PH3} B. B. Back {\it et al.} (PHOBOS), Phys. Rev. Lett.  {\bf 87}, 102303 (2001)

\bibitem{PH4} B. B. Back {\it et al.} (PHOBOS),	Phys. Rev. C {\bf 65}, 031901R (2002) 

\bibitem{PH5} B. B. Back {\it et al.} (PHOBOS), Phys. Rev. Lett. {\bf 88}, 022302 (2002) 

\bibitem{PH7} B. B. Back {\it et al.} (PHOBOS), Phys. Rev. C {\bf 65}, 061901R (2001)

\bibitem{PH10} B. B. Back {\it et al.} (PHOBOS), Phys. Rev. Lett. {\bf 91}, 052303 (2003) 

\bibitem{PH15} B. B. Back {\it et al.} (PHOBOS), Phys. Rev. Lett. {\bf 93}, 082301 (2004)

\bibitem{PH17} B. B. Back {\it et al.} (PHOBOS), Phys. Rev. C {\bf 70}, 021902(R) (2004)

\bibitem{PH23} B. B. Back {\it et al.} (PHOBOS), Phys. Rev. C {\bf 72},031901 (2005)

\bibitem{PH25} B. B. Back {\it et al.} (PHOBOS), Phys. Rev. C {\bf 74},021901 (2006)

\bibitem{PH29} B. B. Back {\it et al.} (PHOBOS), Phys. Rev. C {\bf 74},021902 (2006)

\bibitem{PH34} B. Alver {\it et al.} (PHOBOS), Phys. Rev. Lett. {\bf 102},142301 (2009)

\bibitem{PH24} B. B. Back {\it et al.} (PHOBOS), Nucl. Phys. {\bf A 757}, 28 (2005)

\bibitem{BRAHMSmult} I. Arsene {\it et al.}(BRAHMS), Nucl.Phys. {\bf A757}, 1 (2005)

\bibitem{STARmult} Adler C., {\it et al.} (STAR), Phys. Rev. Lett., {\bf 87} 112303 (2001) 

\bibitem{PHENIXmult} Adcox K., {\it et al.} (PHENIX), Phys. Rev. Lett. {\bf 86}, 3500 (2001)

\bibitem{refA} J. E. Elias {\it et al.}, [E178], Phys. Rev {\bf D22}, 13 (1980)

\bibitem{refB} D. H. Brick {\it et al.}, [E565/570], Phys. Rev. {\bf D39}, 2484 (1989)

\bibitem{refC} C DeMarzo {\it et al.}, [NA5], Phys. Rev. {\bf D26}, 1019 (1982)

\bibitem{refD} S. Fredriksson {\it et al.}, Phys. Rep. {\bf 144}, 187 (1987)


\bibitem{NA49}  S. V. Afanasiev {\it et al.}, Phys. Rev. C {\bf 66}, 054902 (2002);
			T. Anticic {\it et al.}, Phys. Rev. C {\bf 69}, 024902 (2004)

\bibitem{Klay} J.L.Klay {\it et al.} (E895), Phys. Rev. {\bf C68}, 054905 (2003) 


\bibitem{PaddleCounters} R. Bindel {\it et al.}, Nucl. Instr. Meth. A {\bf 474}, 38 (2001) 

\bibitem{Hollis_Bari} R.S. Hollis {\it et al.}, Proceedings of  "Intl. Wkshp. Part. Mult. Rel. HI Coll." 17–19 June 2004, Bari, Italy, 
		J.Phys Conf. Ser. {\bf 5}, 46 (2005)

\bibitem{Sagerer_thesis} J. F. Sagerer, Ph.D. Thesis, University of Illinois at Chicago (2008) 

\bibitem{Bialas} A. Bialas, M. Bleszynskiand, and W.Czyz, Nucl. Phys. {\bf B111}, 461 (1976)
 
\bibitem{Busza} W. Busza {\it et al.} (E178) Phys. Rev. Lett. {\bf 34}, 836 (1975)


\bibitem{Glauber}	We use the HIJING Glauber model, which is a Monte Carlo method, to estimate the cross section integrals. This method is generally accepted to be more accurate than analytical methods.

\bibitem{RN}R. Nouicer {\it et al.} (PHOBOS), Nucl. Instr. Meth. {\bf A461} (2001)143.  

\bibitem{Ny}E. Nygard {\it et al.}, Nucl. Instr. Meth. {\bf A301}(1991)506.

\bibitem{To}O. Toker {\it et al.}, Nucl. Instr. Meth. {\bf A340}(1994)572.

\bibitem{Calibration} E. Griesmayer {\it et al.}, 
				IEEE Trans. Nucl. Sci. {\bf NS-48}, 1565 (2001) 

\bibitem{Wozniak} K. Wozniak {\it et al.} (PHOBOS), Nucl. Instr. Meth. {\bf A566}, 185 (2006)

\bibitem{Garcia} E. Garcia {\it et al.}, Nucl. Instr. Meth. {\bf A570}, 536 (2007)


\bibitem{HIJING} X.-N. Wang and M. Gyulassy, Phys. Rev. D {\bf 44},3501 (1992); {\bf 45}, 844 (1992); M. Gyulassy and X.-N. Wang, Comp. Phys. Comm. {\bf 83}, 307 (1994); X.-N. Wang, Phys. Rep. {\bf 280}, 287 (1997); Nucl. Phys. {\bf A661}, 609c (1999)


\bibitem{Geant} GEANT Detector Description and Simulation Tool v. 3.21, CERN Program Library, W5013 (1994)

\bibitem{Venus} K. Werner, Phys. Rep. {\bf 232}, 87 (1993)

\bibitem{RQMD}  H. Sorge, Phys. Rev. C {\bf 52}, 3291 (1995)

\bibitem{Reuter_thesis} M. Reuter, Ph. D. Thesis, University of Illinois at Chicago, (2001)


\bibitem{PDB} K. Hagiwara {\it et al.}, 
				Phys. Rev. D {\bf 66}, 010001 (2002)

\bibitem{NPB129_365} W. Thom\'{e} {\it et al.} (ISR), 
				Nucl. Phys. B {\bf 129}, 365 (1977).


\bibitem{Alner} G. J. Alner {\it et al.}, 
				Z. Phys. C {\bf 33}, 1 (1986)
\bibitem{ALICE} K. Aamodt {\it et al.} (ALICE), arXiv:1004:3034 (2010)
\bibitem{PRD41_2330} F. Abe {\it et al.}, 
				Phys. Rev. D {\bf 41}, 2330 (1990).
\bibitem{Morse} W.D.Morse {\it et al.}, Phys. Rev. D {\bf 15}, 66 (1977)

\bibitem{CMS}   V. Khachatryan {\it et al.} (CMS), JHEP02, 41 (2010)

\bibitem{CMS_7TeV} V. Khachatryan {\it et al.} (CMS), Phys. Rev. Lett. {\bf 105}, 022002 (2010)







\bibitem{AGS} J. Klay {\it et al.}, Phys. Rev. C {\bf 68}, 054905 (2003);
			L. Ahle {\it et al.}, Phys. Rev. C {\bf 57}, R466 (1998);
			B. B. Back {\it et al.}, Phys. Rev. C {\bf 66}, 054901 (2002);
			J. Dunlop, Ph.D. thesis, Massachusetts Institute of Technology (1999);
			L. Ahle {\it et al.}, Phys, Lett. B {\bf 490}, 53 (2000) - the width $\sigma=\sqrt{\ln(s/4m^2)/2}$ of the $dN/dy$ distribution was estimated from Landau hydrodynamics, see  {\it e.g.} P. Carrouthers and Ming Duong-van, Phys. Rev. D {\bf 8}, 859 (1973)



				Phys. Lett. B {\bf 523}, 79 (2001)



\bibitem{Bari} B.B.Back, Proceedings of  "Intl. Wkshp. Part. Mult. Rel. HI Coll." 17–19 June 2004, Bari, Italy, 
		J.Phys Conf. Ser. {\bf 5}, 1 (2005)

\bibitem{Heiselberg} H. Heiselberg, Phys. Rep. {\bf 351}, 161 (2001)
 








\bibitem{Benecke} J. Benecke {\it et al.}, Phys. Rev. {\bf 188}, 2159 (1969)


\bibitem{PH19} B.B. Back {\it et al.} (PHOBOS), Phys. Rev. Lett. {\bf 94}, 122303 (2005)












\bibitem{refE} I. Otterlund {\it et al.}, Nucl. Phys. {\bf B142}, 445 (1978)


\end{references}
\end{document}